\documentclass[12pt]{article}
\usepackage{amsmath}
\usepackage{amssymb}
\usepackage{amsfonts}
\usepackage{latexsym}
\usepackage{epsfig}
\input epsf.sty

\begin{document}
\newcommand\field[1]{{\ensuremath{\mathbb{{#1}}}}}

\newcommand{\UU}{\mathcal{U}}
\newcommand{\RR}{\field{R}}
\newcommand{\TT}{\field{T}}
\newcommand{\sR}{\mathcal{R}}
\newcommand{\p}{\partial}
\newcommand\nab{{\nabla}}
\newcommand{\OO}{\mathcal{O}}
\newcommand{\HH}{\mathcal{H}}

\def\ov{\over}
\def\le{\left}
\def\ri{\right}
\def\ha{{1\over 2}}
\def\lam{{\lambda}}
\def\Lam{{\Lambda}}
\def\al{{\alpha}}
\def\ket#1{|#1\rangle}
\def\bra#1{\langle#1|}
\def\vev#1{\langle#1\rangle}
\def\det{{\rm det}}
\def\tr{{\rm tr}}
\def\Tr{{\rm Tr}}
\def\NN{{\cal N}}
\def\th{{\theta}}
\def\De{{\Delta}}

\def\Om{{\Omega}}
\def \th{{\theta}}

\def \lam {\lambda}
\def \om {\omega}
\def \ra {\rightarrow}
\def \ga {\gamma}
\def\sig{{\sigma}}
\def\ep{{\epsilon}}
\def\apr{{\alpha'}}
\def\LL{{\cal L}}
\def\tir{{\tilde r}}
\def\Ga{{\Gamma}}

\def\LL{{\cal L}}
\def\NN{{\cal N}}
\def\CC{{\cal C}}

\newcommand{\be}{\begin{equation}}
\newcommand{\ee}{\end{equation}}
\newcommand{\bea}{\begin{eqnarray}}
\newcommand{\eea}{\end{eqnarray}}
\newcommand{\nn}{\nonumber\\}

\newcommand{\bln}{\begin{align}}
\newcommand{\eln}{\end{align}}
\newcommand{\bst}{\begin{split}}
\newcommand{\est}{\end{split}}
\newcommand{\bi}{\begin{itemize}}
\newcommand{\ei}{\end{itemize}}
\newcommand{\ben}{\begin{enumerate}}
\newcommand{\een}{\end{enumerate}}

\title{\vbox{\baselineskip12pt
\hbox{\normalsize MIT-CTP/4006}
\hbox{}
\hbox{}
}
Graviton and Scalar Two-Point Functions\\in a CDL Background for General Dimensions}

\author{
  Daniel S. Park \thanks{email: whizpark@mit.edu} \\
	\em Center for Theoretical Physics, MIT\\
        \em Cambridge, Massachusetts 02139, USA\\
}

\date{March 10, 2009}

\maketitle

\begin{abstract}
We compute the two-point functions of the scalar and graviton in a
Coleman-De Luccia type instanton background in general dimensions.
These are analytically continued to Lorentzian signature.
We write the correlator in a form convenient for
examining the ``holographic" properties
of this background inspired by the work of Freivogel, Sekino, Susskind and Yeh(FSSY).
Based on this, we speculate on what kind of boundary theory
we would have on this background 
if we assume that there exists a holographic duality.

\end{abstract}

\newpage

\section{Introduction}

The $AdS/CFT$ correspondence \cite{AdSCFT} has shed new light on how to
think about gauge theories and gravity theories in general.
It is natural to try to find such a way of understanding
gravity with a background we live in, namely, in deSitter space.
There have been many ideas put forth on how to think about such matters
\cite{Strominger:2001pn, Dyson:2002nt, Maldacena:2002vr, Freivogel:2006xu, Susskind:2007pv, Bousso:2008as, Garriga:2008ks}.
The current view expressed in the literature is that
since a boundary theory of such a correspondence would have to lie at
timelike infinity, one would have to take into account
bubble nucleation for realistic theories.
The argument is that if we have finite probability for bubble
nucleation(which is very possible by semi-classical arguments
such as \cite{ColemanVac},)
for any path we take to future timelike infinity,
we must encounter some kind of bubble nucleation along the way.

\begin{figure}[!ht]
\leavevmode
\begin{center}
\epsfysize=10cm  
\epsfbox{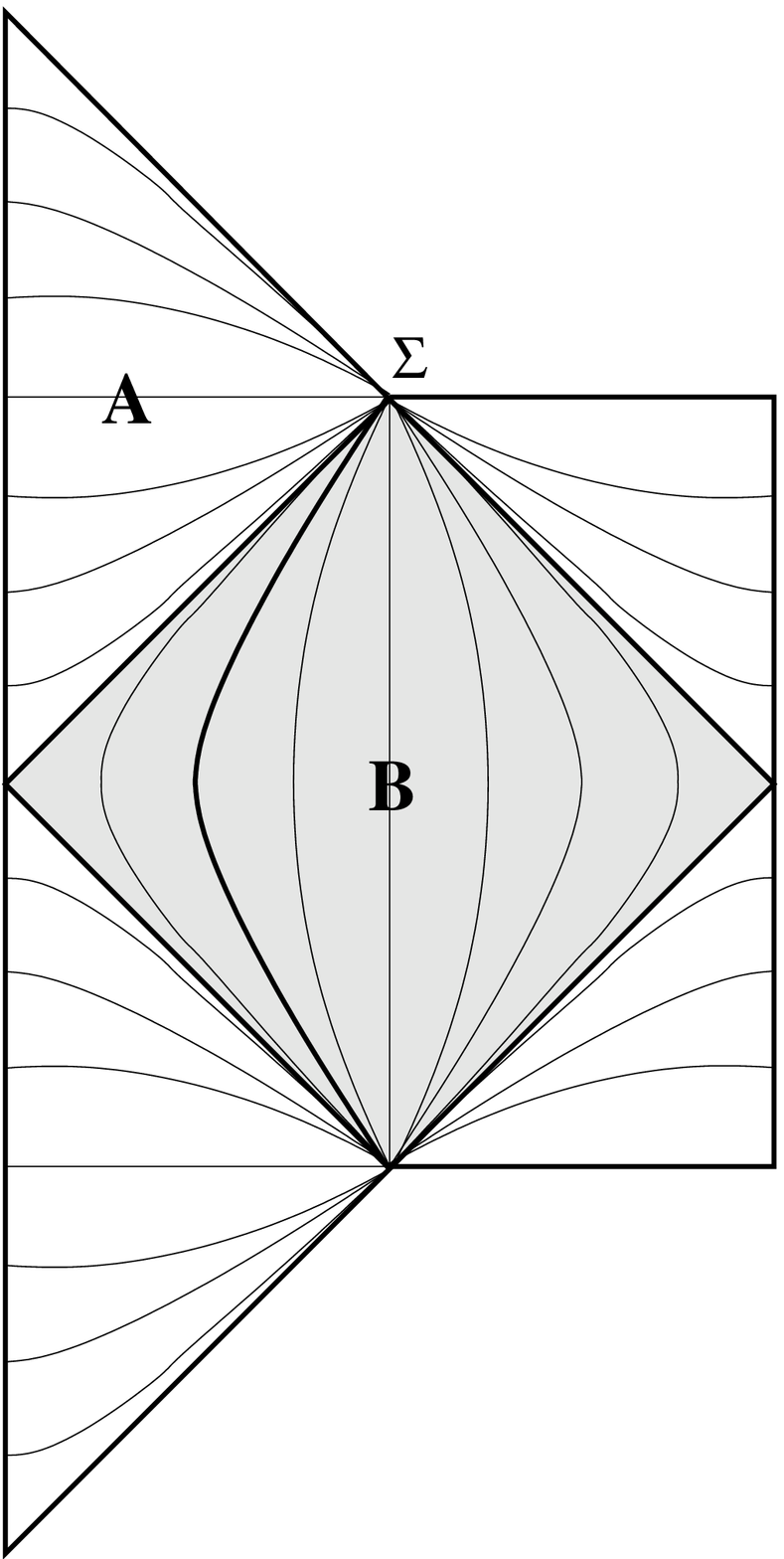}
\end{center}
\caption{\small The Penrose diagram for the Coleman-De Luccia instanton.
The bold curve in the grey region is the bubble wall. The space is flat
to the left of the wall, and deSitter to the right side of the wall.
}
\label{BubblePD2}
\end{figure}

Thus it is natural to consider quantum gravity in
backgrounds with bubble nucleation, for example that
described by a Coleman-De Luccia(CDL) instanton \cite{Coleman:1980aw}.
The Penrose diagram of this instanton (the `bounce' as they put it)
is shown in figure \ref{BubblePD2}.
If we consider the timelike flat region(region A) of this background in $D$ dimensions,
it has a well defined spatial infinity at $\Sigma$, which is a $S^{D-2}$.

Freivogel, Sekino, Susskind, and Yeh proposed that
there may well be some kind of holographic correspondence between the
bulk theory in region A and its boundary $\Sigma$ in \cite{Freivogel:2006xu}.
A further interpretation of the consequences of this calculation was
pursued in \cite{Susskind:2007pv}.
In these papers, they have proposed that in 4 dimensions,
the holographic dual living at $\Sigma$ corresponding
to the bulk gravity theory would be a Liouville theory.
Furthermore, they have identified the conformal time coordinate
with the Liouville field on the boundary.
If this were true, time in the bulk would be emergent
from a Liouville field on the boundary.

This was suggested by writing out the two point functions in
a manner that made the (potentially) holographic structure more
evident and analyzing relevant pieces that showed up in this propagator.
What we will do in this paper is to further carry out
this analysis to higher dimensions.

We will do this by obtaining the two point function of the
transverse traceless graviton and scalar in this background.
This has been done in the past in 4 dimensions\cite{Freivogel:2006xu, Hertog:1999kg},
but we extend the calculation to general dimensions.

This calculation is meaningful in three ways.
First, doing a `holographic expansion' of the propagators in
a general dimensional background gives a clear framework
as to how to do such an expansion in the 4 dimensional case.
Having an explicit $D$ in the expansion helps
organizing the terms in the expression for the propagators.

Also, exploring a potential holographic duality of this kind
in general dimensions turns out to be an interesting topic in itself.
It would be very interesting to see which statements
FSSY have set forth for the 4 dimensional case
still hold in general dimensions.
As we shall see, all of their conjectures
regarding the existence of a holographic duality could be repeated
with slight modification for the general dimensional case.
In fact, the clearer organization of the terms in the holographic
expansion enables us to say a bit more.

Last but not least, we expect gravity in odd and even dimensions
to behave very differently, and it will be interesting to see how
this shows in the propagator.
We will be able to see that if such a correspondence existed,
the boundary theory would in fact be very different for odd and even dimensions.

In this paper,
we will consider a theory with gravity and a single scalar field with a potential that has two
minima in a general dimensional space. We assume the thin wall limit can be applied,
that is, that there is a classical solution of the theory where we have two distinct regions
of space with different cosmological constants seperated by a thin wall.
We will be interested in the case where we have a flat space inside the bubble
and a de Sitter space outside.
We will review this background--the CDL instanton--in section \ref{s:CDL}.

The thin wall introduces a boundary in our space.
We will calculate two point functions
of the transeverse traceless graviton field and the scalar field
inside the timelike region of the thin wall,
and take it to the infinite boundary of the thin wall(which is a $S^{D-2}$,)
and see how it behaves.

We will do this for the graviton two point function in section \ref{s:grav1}.
We will follow the standard procedure of
first calculating the two point function of the graviton
on the Euclidean instanton,
and then analytically continuing it to Lorentzian signature \cite{Gratton:1999ya}.
After that, we will write this as a sum of
massive tensor field propagators in $H^{D-1}$.
We will not be interested in the ordinary part of the propagator,
but the part of the propagator that arises due to the existence of the wall.

As pointed out in \cite{Hertog:1999kg, Hawking:2000ee}
the propagator obtained would still have
some residual gauge freedom we have to project out.
We will explain in section \ref{ss:gcci} the
`naive way' of projecting out those degrees of freedom.
We will obtain the propagator
after this projection in section \ref{ss:largel2}.
Finally, in section \ref{ss:doublepoles} we will point out
the subtlety overlooked by the method of projection employed in
section \ref{ss:gcci} and present the final
gauge-fixed two point function.

We will summarize the graviton two point function
and examine important features we see in it in
section \ref{s:obs}. 

A similar calculation for the scalar is done in section \ref{s:scalar}.
The final `holographic expansion' for the scalar is written out in
section \ref{ss:massless} when the scalar is massless and
in section \ref{ss:accdp} for the general case.
A major difference of the scalar case with the graviton case is pointed out in
the latter section as well.

Finally, we will interpret the results in section \ref{s:sum}.
In this section, we try to guess what the theory living on the
boundary $S^{D-2}$ would look like if we assume a holographic correspondence,
as there are a number of interesting proposals we could make
just from the `holographic expansion' of the propagators.

We first propose the conformal structure of the theory living on the boundary,
and argue that it is highly possible that it contains gravity.
We also propose a possible holographic correspondence between fields
in the bulk with operators on the boundary.
We see how the tunable dimensionful parameter of the theory,
namely the wall position, plays a role in this correspondence.
We pay special attention to the operators whose conformal dimensions
depend on the wall position.
We mention that the number of these are finite in even dimensions,
while they are infinite in odd dimensions.
Based on the behavior of these operators,
we note that in odd dimensions, some kind of phase transition
seems to happen as the dimensions of infinitely many operators
become complex at a critical position of the wall.

As one of the objectives of the paper is to present a thorough
description of the calculation procedure of the graviton two point function,
there are many technical details that might be uninteresting to some readers.
I believe reading section \ref{s:CDL} for understanding the instanton background
we are working in, and sections \ref{s:obs} and \ref{s:sum} for seeing the results
and implications of the calculation would be enough
for those who wish to skip such details.

\section{The Background} \label{s:CDL}

We consider a theory with gravity and a single scalar field $\phi$ with a potential $V(\phi)$,
in a $D$ dimensional space. We assume $D$ is even.

By setting $\kappa = 8\pi G = 1/2$, our Lagrangian would look like,
\be
 S = \ha \int d^D x \sqrt{-g} (- g^{\mu\nu} \nab_\mu \phi \nab_\nu \phi - 2V(\phi) + 2R)
\label{Lag}
\ee
where we use the sign convention, $(-++\cdots+)$. We wish to obtain a classical
solution for this action, and expand around that background.
Now $V$ in general could be anything,
but we are interested in the situation of
tunneling between ground states with positive and zero
scalar vacuum expectation values.
Hence we assume that $V$ has two local minima,
each at $\phi_+ , \phi_-$, with $V(\phi_+) > V(\phi_-)=0$.

We may follow the course of Euclideanizing the action, solving for it,
then analytically continuing it.
Also, we assume an $O(D-1)$ symmetry of the solution for Euclidean metric,
and furthermore assume that the classical solution for $\phi$ is only
a function of the radial coordinate.
This symmetry may not exist for all classical solutions, but we are not
interested in cases that do not respect this symmetry.
So we may begin by setting the metric of the
$D$ dimensional Euclidean manifold as,
\begin{align}
	ds^2 &= dt^2 + a(t)^2 (d \theta^2 + \sin^2 \theta d \Omega_{D-2}^2 )
\end{align}

Then we obtain the classical solution by solving,
\begin{align}
\label{BEOM1}
 & \ddot{\phi} + (D-1) {\dot{a} \over a} \dot{\phi} = {dV \over d \phi}  \\
\label{BEOM2}
 & \dot{a}^2 = 1 + {a^2 \over (D-1)(D-2)} (\ha \dot{\phi}^2 - V(\phi))
\end{align}
with boundary conditions,
\be
 \dot{a} = 1 ~(t=0), ~~~~ \dot{a} = -1 ~(t=t_1), ~~~~ \dot{\phi}=0~(t=0, t=t_1)
\ee
where the dot implies differentiation with respect to $t$.
These boundary conditions correspond to the situation where we have $\phi$
settled safely at their minima for coordinates, $t=0, t_1$.
We are particularly interested in the thin wall limit,
where we may approximate,
$\phi=\phi_-$ for $t<t_0$ and $\phi=\phi_+$ for $t>t_0$.
This would correspond to a `bubble' with different cosmological constants on
either side.
The metric would yield as that of a maximally symmetric space
with given cosmological constants.
Since we have assumed that $V(\phi_+) > V(\phi_-)=0$,
we would have a flat space in the region $t<t_0$
and (Euclidean) de Sitter space in $t>t_0$.

Before turning back to Lorentzian signature,
we wish to convert to conformal
coordinates, that is, coordinates such that,
\be
 dX = dt / a(t)
\ee

Then we may write the metric as,
\begin{align}
 ds^2 &= a^2 (X) (dX^2 + d \theta^2 + \sin^2 \theta d \Omega_{D-2}^2 )
\end{align}

The metric for the CDL instanton 
can be written in this coordinate system as,
\begin{align}
 a(X) = \begin{cases}
      \frac{e^{X-X_0 }}{\cosh X_0 } & (X < X_0) \\
      \frac{1}{\cosh X} & (X > X_0) \\
\end{cases}
\label{aX}	
\end{align}

The analytic continuation required to obtain the timelike flat region is,
\begin{align}
	X = T + i \pi /2,~~~~ i a(T) = a(T + i \pi /2),~~~~ \theta \rightarrow i R ~(\mu \rightarrow i l)
\label{ancon}
\end{align}
which sends slices of $(D-1)$-spheres to slices of $(D-1)$-hyperbolic spaces.
The $\mu$ is the geodesic distance on the sphere, where $l$ is the geodesic distance on the
hyperbolic space. This yields the metric,
\begin{align}
	ds^2 = a(T)^2 ( - dT^2 + dR^2 + \sinh^2 R d \Omega_{D-2} ^2 ) = a(T)^2 (-dT^2 + d H_{D-1} ^2 )
\end{align}
where now,
\be
 a(T) = \left({e^{-X_0} \ov \cosh X_0}\right) e^T
\ee
which provides the metric for the timelike region inside the bubble.
$dH_n^2$ denotes the metric for the
$n$-dimensional hyperbolic space.
Note that we have a well defined spatial infinity in this region, namely at
$R \rightarrow \infty$ which is an $S^{D-2}$.
With respect to figure \ref{BubblePD2}, this metric describes region A.
The thin curves inside this region denotes constant $T$ slices which are $H^{D-1}$s.
$\Sigma$ is at $R \rightarrow \infty$. We will be obtaining the graviton and
scalar propagator between two points in this region and taking it to the boundary
$\Sigma$.

It will prove convenient to use Poincar\'e coordinates to describe
the hyperbolic slices, in which case we obtain the metric,
\be
 ds^2 = a(T)^2 (-dT^2 + { dz^2 + dx_1^2 + \cdots + dx_{D-2}^2 \ov z^2} )
\ee
In these coordinates, $\Sigma$ lies at $z \rightarrow 0$.

For our purposes we are not interested in the spacelike region of the CDL background,
but for the sake of the completeness in the argument,
the continuation that yields the metric for this region is,
\begin{align}
	\theta \rightarrow i t' + \pi/2
\end{align}
which results in the metric,
\begin{align}
	ds^2 &= a^2 (X) (dX^2 - d t'^2 + \cosh^2 t' d \Omega_{D-2}^2 )
\end{align}
This describes region B of figure \ref{BubblePD2},
where the thin curves inside the region denotes constant $X$ slices,
and the thick bubble wall is at $X=X_0$.

These two regions are patched together at $T=-\infty$ and $X=-\infty$,
which is the thick line in figure \ref{BubblePD2} that divides
region A and B.

\section{The Transverse-Traceless Graviton \\ Propagator} \label{s:grav1}

\subsection{The Equation of Motion}

We first calculate the transverse-traceless tensor
propagator on the Euclidean manifold and writing it out in a form that has
a natural analytic continuation. After that, we will do the analytic continuation
(\ref{ancon}) to the timelike region inside the bubble of our CDL background
and obtain the desired propagator respectively.

Taking the metric for a unit $S^{D-1}$ to be $\tilde{g}_{ij}$,
the whole background metric can be written as,
\begin{align}
 g_{\mu\nu}=
  \begin{pmatrix}
   a(X)^2 & 0\\
   0 & a(X)^2 \tilde{g}_{ij}
  \end{pmatrix}
\end{align}

Also, for convenience, we define,
\begin{align}
 N \equiv \frac{D-2}{2}
\end{align}

We write the metric as,
\begin{align}
  g_{\mu \nu} + \delta g_{\mu \nu}
\end{align}
where $g_{\mu \nu}$ is the background metric.

We are interested in the $O(D-1)$, gauge invariant perturbation,
\footnote{The argument that these perturbations are gauge invariant are presented
in various literature, for example in \cite{Hawking:2000ee}. We will later point out
a subtlety that arises in $H^{D-1}$, namely that certain modes we have to consider
turn out to depend on gauge. We will address these issues in section \ref{s:gauge}.}
\begin{align}
	\delta g_{\mu\nu}=
		\begin{pmatrix}
		0 & 0\\
		0 & a(X)^2 h_{ij}
		\end{pmatrix}
\end{align}
where $h_{ij}$ is transverse, traceless on $S_{D-1}$, that is,
\begin{align}
	\tilde{\nabla}^i h_{ij} = 0,~~~~ h_i^i = 0
\end{align}
where $\tilde{\nabla}$ is the covariant derivative with respect to the
metric $\tilde{g}_{ij}$.
(We will use lowercase greek letters to denote coordinates in $D$ dimensions,
and use letters from the english alphabets
to denote coordinates in its $(D-1)$ slices,
be it Euclidean or Lorentzian.)
It turns out that,
\begin{align}
	\nabla^\mu \delta g_{\mu\nu} = 0,~~~~ \delta g_\mu^\mu = 0
\end{align}
where $\nabla$ is the covariant derivative with respect to the
metric $g_{ij}$.
Hence $\delta g_{\mu\nu}$ is transverse traceless with respect to
$g_{ij}$ also. Defining,
\begin{align}
	\tilde{h}_{ij} = a^{(D-2)/2}(X) h_{ij}
\end{align}
the relevant part of the action concerning $\tilde{h}_{ij}$ is,
\begin{align}
	S = \frac{1}{2} \int dX d \Omega_{D-1} \sqrt{\tilde{g}} \tilde{h}_{ij} [-\partial_X^2 +U(X)+2- \widetilde{\Box} ]
            \tilde{h}^{ij}
\end{align}
where $\widetilde{\Box} = \tilde{\nabla}^i\tilde{\nabla}_i$,
$\tilde{g} = \det \tilde{g}_{ij}$ and $U$ is defined as,
\be
U \equiv (a^N)'' / a^N
\label{defU}
\ee
where $f'$ denotes $d f / dX$, and $a(X)$ is given by (\ref{aX}).

Hence if we define,
\begin{align}
 \hat{G}^{ij}_{~~i'j'} (X_1, X_2 , \Omega_1, \Omega_2 ) = a^N (X_1 ) a^N (X_2 ) <h^{ij} (X_1, \Omega_1)
 h_{i'j'} (X_2, \Omega_2)>
\end{align}
this satisfies,
\begin{align}
\begin{split}
 [-\partial_{X_1}^2 +U(X_1) +2 - \widetilde{\Box}_1]\hat{G}^{ij}_{~~i'j'} &(X_1, X_2 , \Omega_1, \Omega_2 ) \\
 &= \frac{1}{\sqrt{\tilde{g}}} \delta(X_1 - X_2 ) \delta^{ij}_{~~i'j'} (\Omega_1 , \Omega_2 )
\end{split}
\label{Geq}
\end{align}
where $\delta^{ij}_{~~i'j'} (\Omega_1 , \Omega_2 )$ is the normalized projection operator
onto transverse traceless tensors on $S^{D-1}$.
The subscript 1 implies differentiation with respect to the coordinates, $(X_1, \Om_1)$.
It's worth reminding ourselves again that we are working on a Euclidean manifold.

\subsection{Decomposition}

Due to the $O(D-1)$ symmetry,
the Green's function $G^{ij}_{~~i'j'}$ has to
respect all the isometries of the $(D-1)$ sphere with respect to
the $S^{D-1}$ coordinates of the two points involved,
i.e. it should be a maximally symmetric bitensor
with respect to its $S^{D-1}$ coordinates.
Therefore it should be possible the write it as,
\begin{align}
 \hat{G}^{ij}_{~~i'j'} (X, X', \mu)
\end{align}
where $\mu(\Omega_1 , \Omega_2 )$ is the geodesic distance between
the two points $\Omega_1 , \Omega_2$
(see \cite{Allen:1985wd} for further discussion.)

The solution for the equation (\ref{Geq}) can be written as,
\begin{align}
 \hat{G}^{ij}_{~~i'j'} (X, X', \mu) = \sum^{+i \infty}_{ p = (N+2)i } G_p (X, X' ) W^{ij}_{(p)i'j'}(\mu)
\label{Ghateq}
\end{align}
We will define $G_p$ and $W^{ij}_{(p)i'j'}(\mu)$,
and explain the range of the sum soon.

$W^{ij}_{(p)i'j'}(\mu)$ is a maximally symmetric bitensor on $S^{D-1}$
defined by,
\begin{align}
 W^{ij}_{(p)i'j'}(\mu) = \sum_{u} q^{(pu)ij\dagger}(\Omega) q^{(pu)}_{i'j'}(\Omega')
\label{Wdef}
\end{align}
where $q^{(pu)ij}$ are transeverse traceless eigenmodes of
\begin{align}
 \widetilde{\Box} q^{(pu)ij} = (N^2 + 2 + p^2 ) q^{(pu)ij}
\end{align}
where $u$ denotes all the quantum numbers other than $p$ needed to specify the mode $q$.
These modes are normalized so that
\begin{align}
	\int d^{D-1} x \sqrt{\tilde{g}} q^{(pu)ij} q^{(p'u')*}_{ij} = \delta^{pp'} \delta^{uu'}
\label{qnorm}
\end{align}

Note that by definition $W^{ij}_{(p)i'j'}(\mu)$ satisfies,
\begin{align}
 \widetilde{\Box} W^{ij}_{(p)i'j'}(\mu) &= (N^2 + 2 + p^2 ) W^{ij}_{(p)i'j'}(\mu) \label{Weq1}
\end{align}
and is transverse(for all the indices)
and traceless(for each pair $ij$ and $i'j'$) with respect to $\tilde{g}_{ij}$.

Also, on $S^{D-1}$, we get eigenmodes that are regular on the whole sphere
only for the $p$ values,
$p=(N+2)i, (N+3)i, \dots $ (see \cite{Camporesi:1994ga},)
so by completeness of the basis,
\begin{align}
	\sum^{+i \infty}_{ p = (N+2)i } W^{ij}_{(p)i'j'}(\mu(\Omega , \Omega' )) = \delta^{ij}_{~~i'j'} (\Omega , \Omega' ) 
	/\sqrt{\tilde{g}}
\label{Weq2}
\end{align}

We define $G_p$ to be the $X, X'$ dependent function that satisfies,
\begin{align}
	[-\partial^2_X + U(X) - (p^2+N^2)]G_p (X, X' ) = \delta(X-X')
\label{Gpeq}
\end{align}

From equations (\ref{Weq1}), (\ref{Weq2}), and (\ref{Gpeq}), we see that indeed (\ref{Ghateq}) solves (\ref{Geq}).

\subsection{$G_p (X, X')$} \label{subsec:Gp}

In order to obtain $G_p (X, X')$ satisfying (\ref{Gpeq}),
let's first think about $F_k (X)$ which satisfy
\begin{align}
	[-{d^2 \over dX^2 } + U(X)]F_k (X) &= (k^2 + N^2 ) F_k (X)
\end{align}
for a uniform background without any kind of wall.
We think about the cases when,
$ a \propto (1/\cosh X), e^X$ each corresponding to the dS, and flat background.

Then, defining,
\begin{align}
  W \equiv \ln (a^N), \quad w \equiv \ln a 
\end{align}
we get,
\begin{align}
	U (X) = W'^2 + W '' = (N+1)N w'^2 - N \\
	\tilde{U} (X) = W'^2 - W '' = N(N-1) w'^2 + N
\end{align}
where we've used the property,
\begin{align}
	w'^2 - w'' = 1
\end{align}

The equation,
\begin{align}
  [-{d^2 \over dX^2 } + N(N+1) w'^2 - N ] F_k  = (N^2 + k^2) F_k
\label{wave_eq}
\end{align}
can be solved in terms of the hypergeometric function $F(a,b;c;z)$ by,
\begin{align}
	F_{k, dS} \equiv e^{ikX} F (-N,N+1;1-ik;(1- \tanh X)/2)
\end{align}
for dS, and
\begin{align}
	F_{k, Flat} \equiv e^{ikX} 
\end{align}
for flat space, where we took the boundary conditions to be,
\begin{align}
 F_{k, dS} \rightarrow e^{ikX} ~~~&X \rightarrow \infty \\
 F_{k, Flat} \rightarrow e^{ikX} ~~~&X \rightarrow -\infty
\end{align}

Now let's introduce the wall.
If we have different $w'$ for $X>X_0 $ and $X<X_0 $, we get,
\begin{align}
	w'^2 - w'' = 1 + A_0 \delta(X-X_0 )
\end{align}
for,
\begin{align}
	A_0 = e^{X_0} / \cosh X_0
\end{align}
Hence if we define $A \equiv -NA_0 $, the Schr\"odinger equation,
\be
 	[-{d^2 \over dX^2 } + U(X) ] u_k = E_k u_k
\ee
for $U(X)$ defined by (\ref{defU}) for (\ref{aX}) becomes,
\begin{align}
	[-{d^2 \over dX^2 } + N(N+1) w' - N +A \delta (X -X_0 ) ] u_k = E_k u_k
\label{waveeqml}
\end{align}
where $w=-\ln (\cosh X)$ for $X>X_0$ and $w=X+(\text{constant})$ for $X<X_0$.

Since we already know the eigenfunctions in the separate domains
$X>X_0$ and $X<X_0$, the equation can be solved by
finding how these waves scatter off the domain wall.
For unbounded states, we may write,
\begin{align}
u_{1k} &= \begin{cases}
F_{k, L} + \RR F_{-k, L}   & (X < X_0) \\
\TT F_{k, R}   & (X > X_0)
\end{cases}
\\[10pt]
u_{2(-k)} &= \begin{cases}
\TT_r F_{-k, L}   & (X < X_0) \\
F_{-k, R} + \RR_r F_{k,R}   & (X > X_0)
\end{cases}
\end{align}
for $E_k = k^2 +N^2$ where,
\begin{align}
 F_{k, L} = F_{k, Flat},~~~~ F_{k, R} = F_{k, dS}
\end{align}
and $\RR, \TT, \RR_r, \TT_r$ are
scattering coefficients which depend on $k$.

Solving the boundary conditions to obtain the reflection coefficient $\RR$, we obtain,
\begin{align}
\label{rcoeff}
 \RR = 
 - e^{2ikX_0}	\frac{[F_{k, R}' (X_0)-AF_{k, R} (X_0)]-ik F_{k, R} (X_0)}{[F_{k, R}' (X_0)-AF_{k, R} (X_0)]+ik F_{k, R} (X_0)}
 = e^{2ikX_0} \sR
\end{align}
where $\sR$ can be written in terms of hypergeometric functions as, 
\begin{align}
 \sR =
 -{N(1-t)F(-N+1,N+1;1-ik;t) \ov (ik+N)F(-N,N;1-ik;t)}, \qquad t={e^{-X_0} \ov 2 \cosh X_0}
\label{sR}
\end{align}

We note the following properties of $\sR$:

\begin{enumerate}
    \item The poles $i a_n$ of $\sR$ in the upper half plane correspond to bound states.
          They are purely imaginary, and $a_n \leq N$.
    \item Regardless of the value of $X_0$, $k = iN$ is always a pole of $\sR$.
\end{enumerate}

When $N$ is an integer, $\sR$ has the following properties.
\begin{enumerate}
    \item $\sR$ is a rational function with respect to $k$.
    \item $\sR$ has $N$ other poles, which are pure imaginary and lie between $(-iN, iN)$.
    \item In the limit $X_0 \to -\infty$, the poles other than $iN$ tend to $0, i, \cdots, i(N-1)$.
    \item In the limit $X_0 \to \infty$, the poles other than $iN$ tend to $-i, -2i, \cdots, -iN$.
    \item $k = - i N$ is always a zero of $\sR$.
\end{enumerate}

When $N$ is a half integer, $\sR$ exhibits some very interesting properties.
As in the integer case, all poles lie on the imaginary axis below $k=iN$, and has only a finite number of poles in $[-iN,iN]$,
but in the limit $k \rightarrow -i \infty$ the pole structure varies starkly:
\begin{enumerate}
    \item For $X_0 \geq 0$, $\sR(k)$ has an infinite number of poles on the imaginary axis of the lower half plane.
          (Appendix \ref{ap:poles})
    \item For $X_0 < 0$, $\sR(k)$ has only a finite number of poles on the imaginary axis of the lower half plane,
          but has an infinite number of complex poles on the lower half plane for $-\epsilon<X_0<0$
          for some $\epsilon>0$. (Appendix \ref{ap:poles})
\end{enumerate}

Now we are ready to solve (\ref{Gpeq}).
\begin{align}
\label{Gp}
\begin{split}
 G_p &(X, X' ) = \\
 &\frac{1}{\Delta_p} [\Psi_p^r (X) \Psi_p^l (X') \Theta(X-X') + \Psi_p^l (X) \Psi_p^r (X') \Theta(X'-X)]
\end{split}
\end{align}
where $\Psi_p^r (X)$ is the solution to the Schr\"odinger equation (\ref{waveeqml}),
that goes to $e^{ipX}$ as $X \rightarrow \infty$, and
$\Psi_p^l (X)$ the solution that goes to $e^{-ipX}$ as $X \rightarrow -\infty$.
$\Delta_p$ is defined to be the Wronskian of $\Psi_p^r$ and  $\Psi_p^l$.

Since we can write $\Psi_p^r (X)$, $\Psi_p^l (X)$ in terms of $u_k$, namely,
\begin{align}
  \Psi_p^r (X) = u_{1p} (X),~~~~\Psi_p^l (X) = u_{2 (-p)} (X)
\label{Psi}
\end{align}
and for the flat side of the bubble, we get,
\begin{align}
	G_p (X, X' ) = \frac{i}{2p} (e^{ip \delta X} + \RR (p) e^{-ip \bar{X}}) ~~~~(X, X' <X_0)
\label{Gpfinal}
\end{align}
where
\begin{align}
 \delta X &=
\begin{cases}
 X-X' \quad (X>X') \\
 X'-X \quad (X'>X) \\
\end{cases} \\
 \bar{X} &= X+X'
\end{align}
and the reflection coefficient is given by (\ref{rcoeff}).

\subsection{$W^{ij}_{(p)i'j'}(\mu)$}

The calculation of the maximally symmetric bitensor $W^{ij}_{(p)i'j'}(\mu)$ in this section is
carried out follwing the steps of \cite{Allen:1994yb}.

A maximally symmetric bitensor in general can be written as,
\begin{equation}
\begin{split}
	T_{iji'j'} = & t_1 g_{ij} g_{i'j'} + t_2 [n_{i} g_{ji'} n_{j'} + n_{j} g_{ii'} n_{j'} + n_{i} g_{jj'} n_{i'}
	               + n_{j} g_{ij'} n_{i'}]\\
	             & + t_3 [ g_{ii'}g_{jj'}+g_{ji'}g_{ij'}] + t_4 n_{i}n_{j}n_{i'}n_{j'}
	             + t_5 [g_{ij} n_{i'} n_{j'} + n_i n_j g_{i'j'}]
\end{split}
\end{equation}
where $t_i$ are functions of $\mu$, the length of the geodesic that connects $\Omega$ and $\Omega'$.
Here, $n_i (\Omega, \Omega')$, $n_i' (\Omega, \Omega')$ are unit vectors each at $\Omega$ and $\Omega'$ pointing
away from $\Omega'$ and $\Omega$ respectively. $g_i^{j'}$ is the parallel propagator along the geodesic.

By using the tracelessness of $W^{ij}_{(p)i'j'}(\mu)$, we can write,
\begin{equation}
\begin{split}
	W^{ij}_{(p)i'j'}(\mu) = & Q_p w^I (\alpha_p(z)) t^{ij}_{I~i'j'} |_{z = \cos^2(\frac{\mu}{2})}
\label{tsumW}
\end{split}
\end{equation}
where we have defined
\begin{equation}
\begin{split}
	z \equiv \cos^2(\frac{\mu}{2})
\end{split}
\end{equation}
and
\begin{align}
\label{ttensor1}
	t^{ij}_{1~i'j'} &= [g_{ij}-(D-1)n_i n_j][g_{i'j'}-(D-1)n_{i'} n_{j'}] \\
\label{ttensor2}
  t^{ij}_{2~i'j'} &= 4 n_{(i} g_{j) (i'} n_{j')} + 4 n_{i}n_{j}n_{i'}n_{j'}  \\
\label{ttensor3}
  t^{ij}_{3~i'j'} &= g_{ii'} g_{jj'} +g_{ji'}g_{ij'} -  2g_{ij} n_{i'} n_{j'} -2 n_i n_j g_{i'j'} - 2(D-1) n_{i}n_{j}n_{i'}n_{j'} \\
\label{ttensor}
  t^{ij}_{~~i'j'} &= t^{ij}_{1~i'j'} - Nt^{ij}_{2~i'j'} -Nt^{ij}_{3~i'j'}
\end{align}
We refer the reader to appendix \ref{ap:Wgrav} for the explicit
expression for $w^I (\alpha_p(z))$ and $Q_p$.
We also have defined $t^{ij}_{~~i'j'}$ which will come in handy later.

We first obtain $w^I (\alpha_p(z))$ starting from equation (\ref{tsumW})
up to a constant by imposing transverseness and the condition (\ref{Weq1}).
The result is given by equation (\ref{wIp}).

The normalization constant $Q_p$ given by equation (\ref{Q_p})
is obtained by considering the degeneracy of the modes $q^{(pu)}_{ij}$.
More specifically, this is done by contracting $i'j'$ and $ij$ and
taking $\Omega=\Omega'$ in equation (\ref{Wdef})
and integrating over the whole sphere by $\Omega$.
By doing this, from (\ref{tsumW}),
the r.h.s. of the contracted and integrated (\ref{Wdef}) would yield
some numerical constant(which can be obtained from (\ref{wIp})) times $Q_p$
times the volume of the $(D-1)$ sphere.
The l.h.s. of
the contracted and integrated (\ref{Wdef}) would yield
the degeneracy of the modes $q^{(pu)}_{ij}$
with given $p$, due to equation (\ref{qnorm}).

Note that in order for $\alpha_p$ and $\beta_p$
defined by (\ref{defalph}), (\ref{defbet})
to be well defined,
and hence $W^{ij}_{(p)i'j'}(\mu)$ to be well defined on the whole sphere
(for all $0 \leq z \leq 1$,)
$p$ must have the values $p = i(N +2), i (N +3 ), \dots$.

\subsection{Analytic Continuation}

Since we have obtained $G_p(X,X')$ and $W^{ij}_{(p)i'j'}$ showing
up in equation (\ref{Ghateq}) in the previous two sections, it is straight forward to
write down the hatted propagator for the instanton.
The problem is that we want to analytically continue this to the
time-like Lorentzian region of our background
(namely to carry out equation (\ref{ancon}),)
but this is not a trivial thing to do.

The problem is that we want to think about the propagator
as we take the points concerned to the boundary
sitting at spacelike infinity($l$ : large) of this region.
But by plugging in (\ref{ancon}) to (\ref{Ghateq})
we don't get a convergent sum in this limit.
This is because
\be
 W^{ij}_{(i(N+2+n))i'j'} (il) \sim e^{(2+n)l} t^{ij}_{~~i'j'}
\ee
for large $l$, as can be easily verified by the
asymptotic limit of hypergeometric functions.

\begin{figure}[!b]
\begin{minipage}[b]{0.5\linewidth}
\centering \includegraphics[width=6cm]{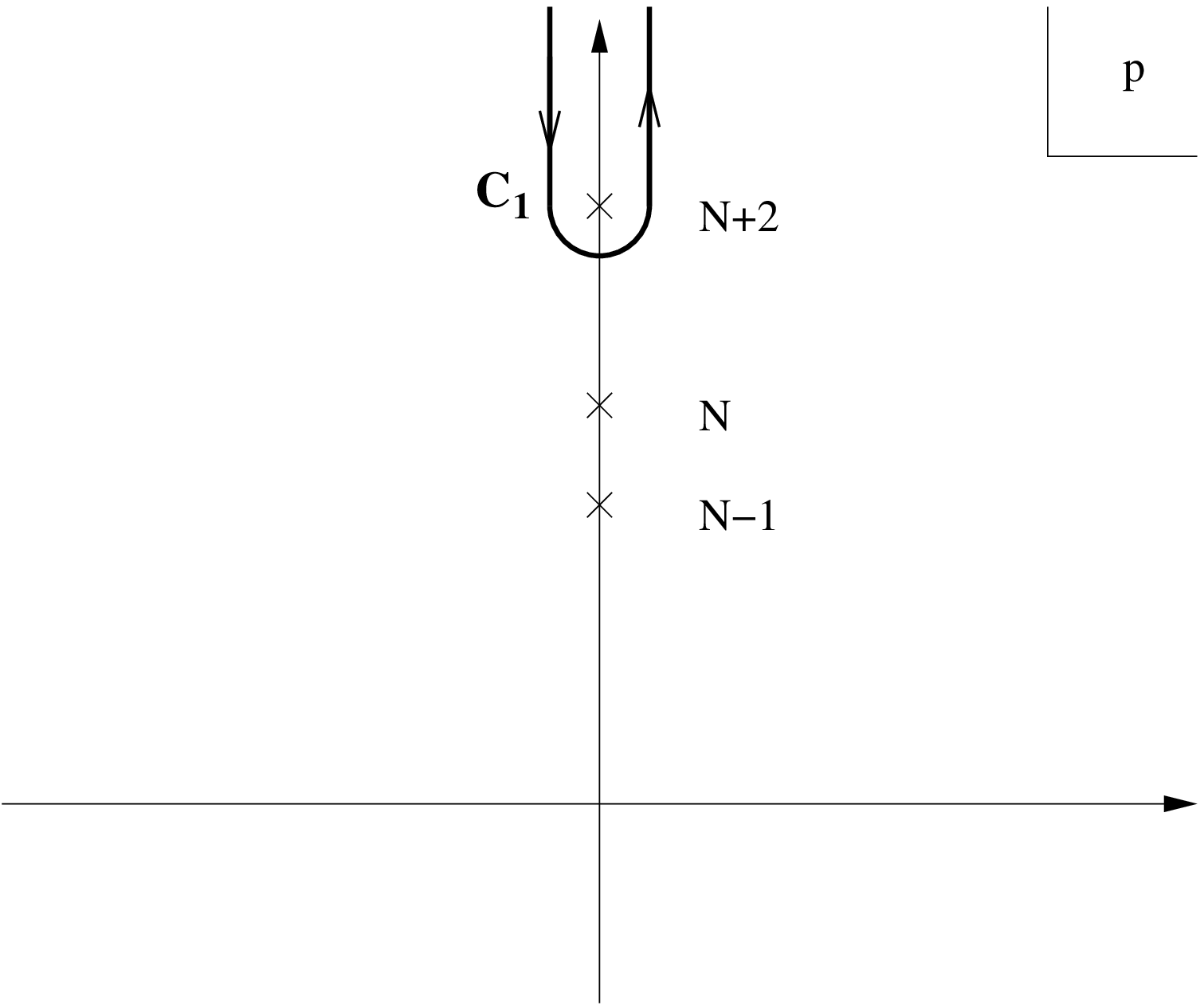}
\caption{\small The contour $C_1$}\label{contour1}
\end{minipage}%
\begin{minipage}[b]{0.5\linewidth}
\centering \includegraphics[width=6cm]{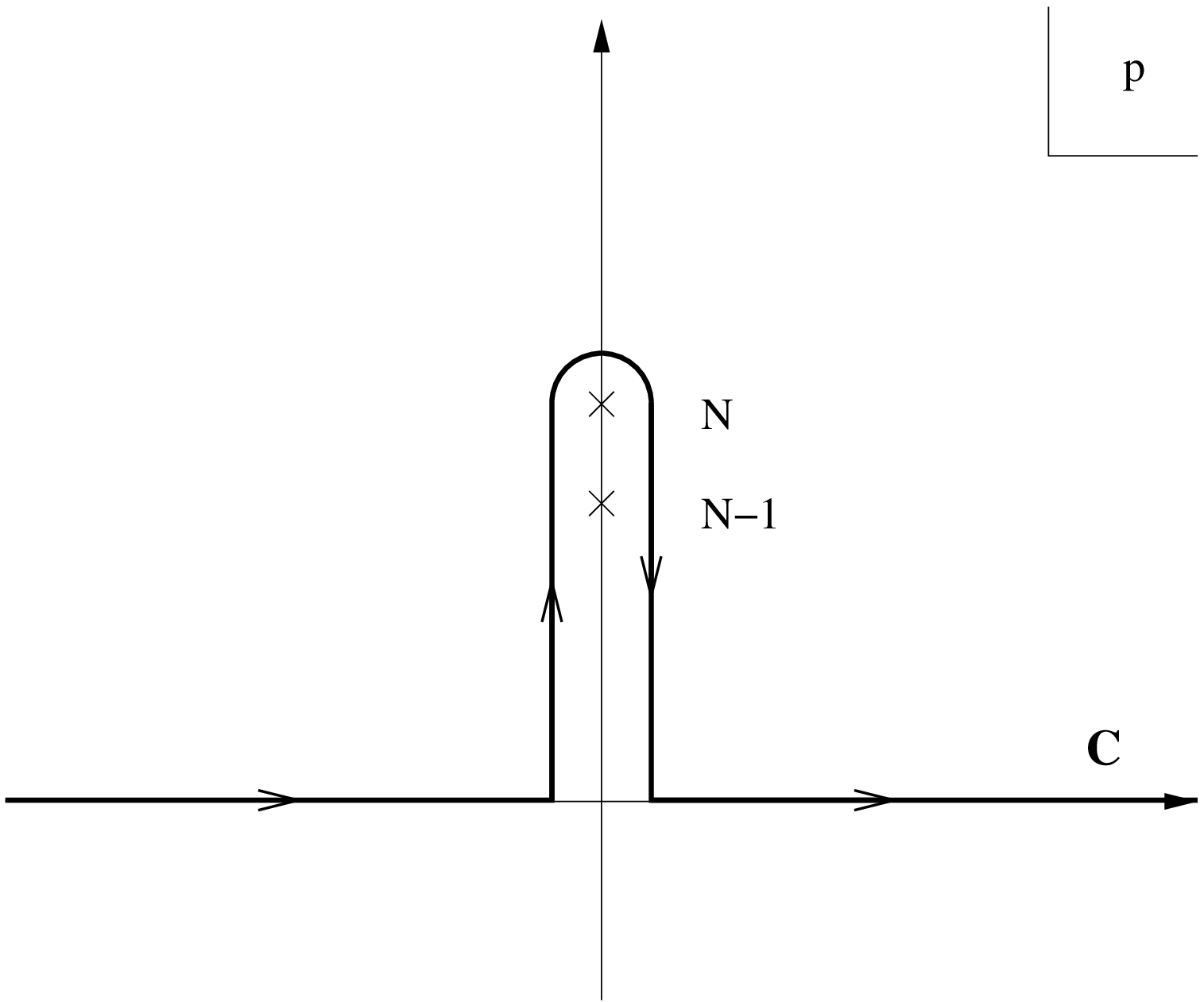}
\caption{\small The contour $C$} \label{contour}
\end{minipage}
\end{figure}

In order to achieve this objective,
we must employ a more sophisticated method
previously utilized by various authors\cite{Freivogel:2006xu, Hertog:1999kg, Gratton:1999ya, Hawking:2000ee}.
The way do this is by expressing the sum (\ref{Ghateq}) as,
\begin{align}
\begin{split}
 \hat{G}^{ij}_{~~i'j'} (X,X', \mu)= \int_{C_1} \frac{dp}{2\pi i} &{\Ga(-ip-N-1)\Ga(ip+N+2) \over (-1)^{-ip-N-2}} \\
                                    & \times G_p (X,X') W^{ij}_{(p)i'j'} (\mu) 
\end{split}
\end{align}
where the contour $C_1$ is defined to be one that comes down from $i \infty$ on the left side of the imaginary
axis of the complex $p$ plane, and pivots around $p=i(N+2)$ to go back to
$i \infty$ by the right side of the imaginary axis.
The $\Ga$ functions pick out the appropriate poles with the desired residues.
This is depicted in figure \ref{contour1}.

Plugging in (\ref{Gpfinal}) to the above expression, we obtain,
\begin{align}
\begin{split}
 \hat{G}^{ij}_{~~i'j'} (X,X', \mu)= \int_{C_1} \frac{dp}{4 \pi p}
  & {\Ga(-ip-N-1)\Ga(ip+N+2) \over (-1)^{-ip-N-2}} \\
  & \times (e^{ip \delta X} + \RR(p) e^{-ip \bar{X}}) W^{ij}_{(p)i'j'} (\mu) 
\end{split}
\end{align}
The first term yields the Green's function for a flat background. Let's focus our attention to the second term,
which we denote by $\hat{G}^{ij~\bar{X}}_{~~i'j'}$.

Now this contour can be safely deformed to
the contour $C$, which we define to run along the real axis of the $p$ plane, with a `jump' over $p=iN$.
This is depicted in figure \ref{contour}.
The contour deformation is justified by the following reasons.

First, by writing the previous equation as,
\begin{align}
 \hat{G}^{ij~\bar{X}}_{~~i'j'} (X,X', \mu)= \int_{C_1} dp (\frac{i}{2p})
  {e^{ip(2X_0 - \bar{X})} \over 1-e^{2 \pi (p-iN)}} \sR(p)  W^{ij}_{(p)i'j'} (\mu) 
\label{Greenprelim}
\end{align}

Since $\bar{X}-2X_0 <0$, the first piece in the integrand decays exponentially
at infinity on the upper half plane, as long as the contour does not pass through $p=in$.
We also know that $F(a,b;c;z) \approx 1+\OO(1/|c|)$ for $c \rightarrow \infty$ so
$\sR(p)$ behaves nicely in this region.

Also, we note that $\alpha_p$ can be written as,
\begin{align}
\begin{split}
 \alpha_p (z) & =F(N+2+ip,N+2-ip;N+5/2;z) \\
              & =\Ga(N+5/2) (z-z^2)^{3/2-N} P^{3/2-N}_{ip-1/2} (1-2z) \\
              & \sim ({ 1 \ov -ip})^{N-1}
\end{split}
\end{align}
for $p \rightarrow i\infty$ and hence $W^{ij}_{(p)i'j'} (\mu) $ also behaves nicely.

Finally, there aren't any poles in the integrand between $iN$ and $i(N+2)$ on the imaginary axis,
(since by (\ref{Q_p}), $W^{ij}_{(p)i'j'} (\mu) =0$ at $p=i(N+1)$)
so we may carry out the deformation as we please. Hence,
\begin{align}
\begin{split}
 \hat{G}^{ij~\bar{X}}_{~~i'j'} (X,X', \mu)= \int_{C} \frac{dp}{4 \pi p} &
  {\Ga(-ip-N-1)\Ga(ip+N+2) \over (-1)^{-ip-N-2}} \\
  & \times \RR(p) e^{-ip \bar{X}} W^{ij}_{(p)i'j'} (\mu) 
\end{split}
\end{align}

Now let's do the analytic continuation,
\begin{align}
  X = T + i \frac{\pi}{2},~~~ \mu = il
\end{align}
Then after pulling out all the trivial constants out in front and sorting out the terms,
the analitically continued propagator piece
$\hat{G}^{ij~\bar{T}}_{~~i'j'} (T, T', l)$ can finally be written as,
\begin{align}
\begin{split}
 \hat{G}^{ij~\bar{T}}_{~~i'j'} (T,T', l) = C_0 &\int_{C} dp 
  \RR e^{-ip \bar{T}} Y^{ij}_{(p)i'j'} (il) \\
 \times & (p^2+(N+1)^2) \Ga(ip+N-1)\Ga(-ip+N-1)
\label{Greenfinal1}
\end{split}
\end{align}
where we have conveniently defined,
\be
Y^{ij}_{(p)i'j'} (il) \equiv {1 \over Q_p} W^{ij}_{(p)i'j'} (il)  = w^I (\alpha_p (z) ) t^{ij}_{I~i'j'}|_{z=\cosh^2 {l \over 2 }}
\ee

\subsection{The Large $l$ Limit}\label{ss:largel}

In this section, we will write out the
`holographic expansion' for the graviton propagator,
i.e. in a form convenient to examine its potential
holographic duality.
In order to do this, it is convenient to invoke the
`generalized Green function's we have defined in
appendix \ref{ap:tpinH}.

We first define,
\begin{align}
 a_p (z) = ({1 \over z})^{{(D+2) \over 2} - ip}
 F (\frac{D+2}{2} -ip, \frac{1}{2} -ip ; 1-2ip ; {1 \over z} )
\end{align}
and then define $G^{ij}_{H~i'j'}$ to be,
\be
  G^{ij}_{H~i'j'} (l,\Delta) = w^I (a_{i(\Delta-N)}) t^{ij}_{I~i'j'} |_{z=\cosh^2 {l \over 2 }}
\label{massprop}
\ee

At large $l$, (or for the Poincar\'e coordinates in $H^{D-1}$, small $z$,)
this behaves as,
\begin{align}
\begin{split}
  G^{ij}_{H~i'j'} & (l,\Delta)  \sim C (\!\Delta\!-\!2N\!)(\!\Delta\!-\!2N\!+1) e^{-\Delta l} t^{ij}_{~~i'j'} +
                                   \OO \left( e^{-(\Delta+2) l} \right) \\
  & \sim C (\!\Delta\!-\!2N\!)(\!\Delta\!-\!2N\!+1) {z^{\Delta} z'^{\Delta} \ov |x-x'|^{2\Delta}}  t^{ij}_{~~i'j'}
     + \OO \left( {z^{\Delta+2} z'^{\Delta+2} \ov |x-x'|^{2\Delta+4}} \right) 
\end{split}
\label{gscaling}
\end{align}
The $\Delta$ dependence of the coefficient of the leading order behavior will prove
to be important.\footnote{Thanks to Leonard Susskind and Yasuhiro Sekino in helping me realize this.}
Also,
\be
  G^{ij}_{H~i'j'} (l,\Delta) \propto G^{ij}_{M~i'J'} (l,\Delta(\Delta-2N))
\ee
for $\Delta > N,~\Delta \neq 2N$, where
$G^{ij}_{M~i'J'}(l,m)$ is the massive transverse traceless propagator
on $H^{D-1}$ with mass $m$. We know from $AdS/CFT$ that this corresponds to
a two point function for a dimension $\Delta$ traceless tensor of the
boundary theory of the $EAdS_{D-1}$ \cite{Polishchuk:1999nh}.

Due to the identity between hypergeometric functions,
\be
 \alpha_p (z) =  {\Ga(N+ {5 \over 2}) \Ga(-2ip)  \over \Ga(N+2-ip) \Ga({1 \over 2}-ip)} a_{-p} (z)
               + {\Ga(N+ {5 \over 2}) \Ga(2ip)  \over \Ga(N+2+ip) \Ga({1 \over 2}+ip)} a_{p} (z)
\ee
so using the linearity of $w^I$, we may write (\ref{Greenfinal1}) as,
\begin{align}
\begin{split}
 &\hat{G}^{ij~\bar{T}}_{~~i'j'} =\\
 &~C_0 \int_{C} dp \RR e^{-ip\bar{T}}
 [{\Ga(-ip) \Ga(ip+N-1) (N+1+ip) \over 2^{-2ip-1/2} (N-ip)(N-1-ip)} w^I (a_{-p}) t^{ij}_{I~i'j'} \\
 & \qquad \qquad \qquad +{\Ga(ip) \Ga(-ip+N-1) (N+1-ip) \over 2^{2ip-1/2} (N+ip)(N-1+ip)} w^I (a_{p}) t^{ij}_{I~i'j'}]
\end{split}
\end{align}
where we have absorbed some overall factors into $C_0$.
We have used the fact that,
\be
 w^I (\alpha_p (z) ) = w^I(c_1 a_{ip} (z) + c_2 a_{-ip} (z)) = c_1 w^I(a_{ip} (z)) + c_2 w^I(a_{-ip} (z))
\ee

This can be re-written as,
\begin{align}
\label{updown0}
\begin{split}
 &\hat{G}^{ij~\bar{T}}_{~~i'j'} =\\
 &~C_0 \int_{C} dp \RR e^{-ip\bar{T}}
 [{\Ga(-ip) \Ga(ip+N-1) (N+1+ip) \over 2^{-2ip-1/2} (N-ip)(N-1-ip)} G^{ij}_{H~i'j'} (l,N+ip) \\
 & \qquad \qquad \qquad +{\Ga(ip) \Ga(-ip+N-1) (N+1-ip) \over 2^{2ip-1/2} (N+ip)(N-1+ip)} G^{ij}_{H~i'j'} (l,N-ip)]
\end{split}
\end{align}
by (\ref{gscaling}). In the large $l$ limit,
\be
 G^{ij}_{H~i'j'} (l,N\pm ip) \sim e^{-(N \pm ip)l} t^{ij}_{~~i'j'}
\ee
Hence in this limit, $C$ for the former term of equation (\ref{updown0}) may be
deformed downward while the latter term may be deformed upward.
This is because the asymptotic behavior
of all the other components in the product
is at worst $\sim e^{ap}$
at $|p| \rightarrow \infty$ on the half-plane concerned
for some fixed number $a$.

\begin{figure}[!b]
\begin{minipage}[b]{0.5\linewidth}
\centering \includegraphics[width=6cm]{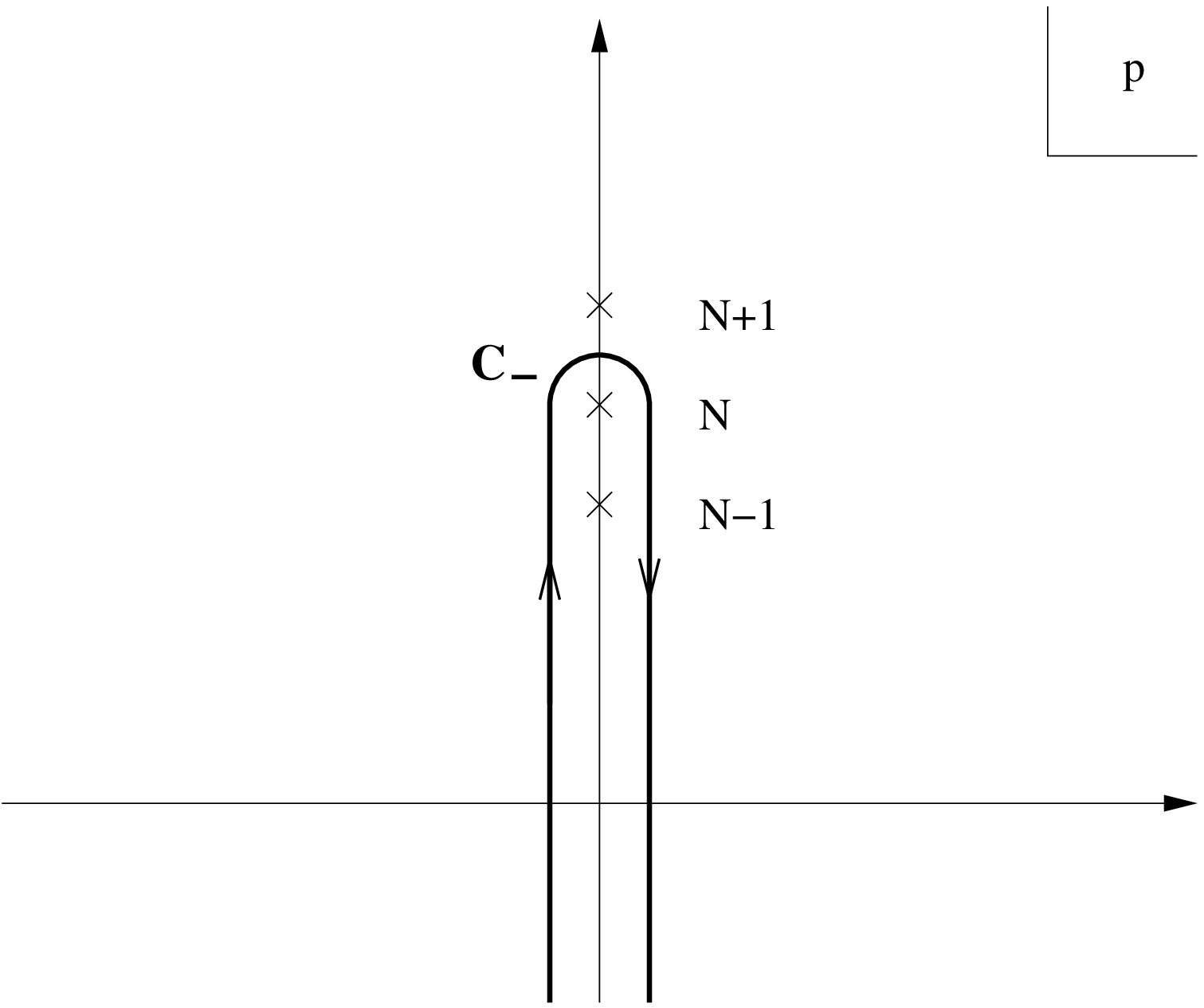}
\caption{\small The contour $C_-$}\label{contourm}
\end{minipage}%
\begin{minipage}[b]{0.5\linewidth}
\centering \includegraphics[width=6cm]{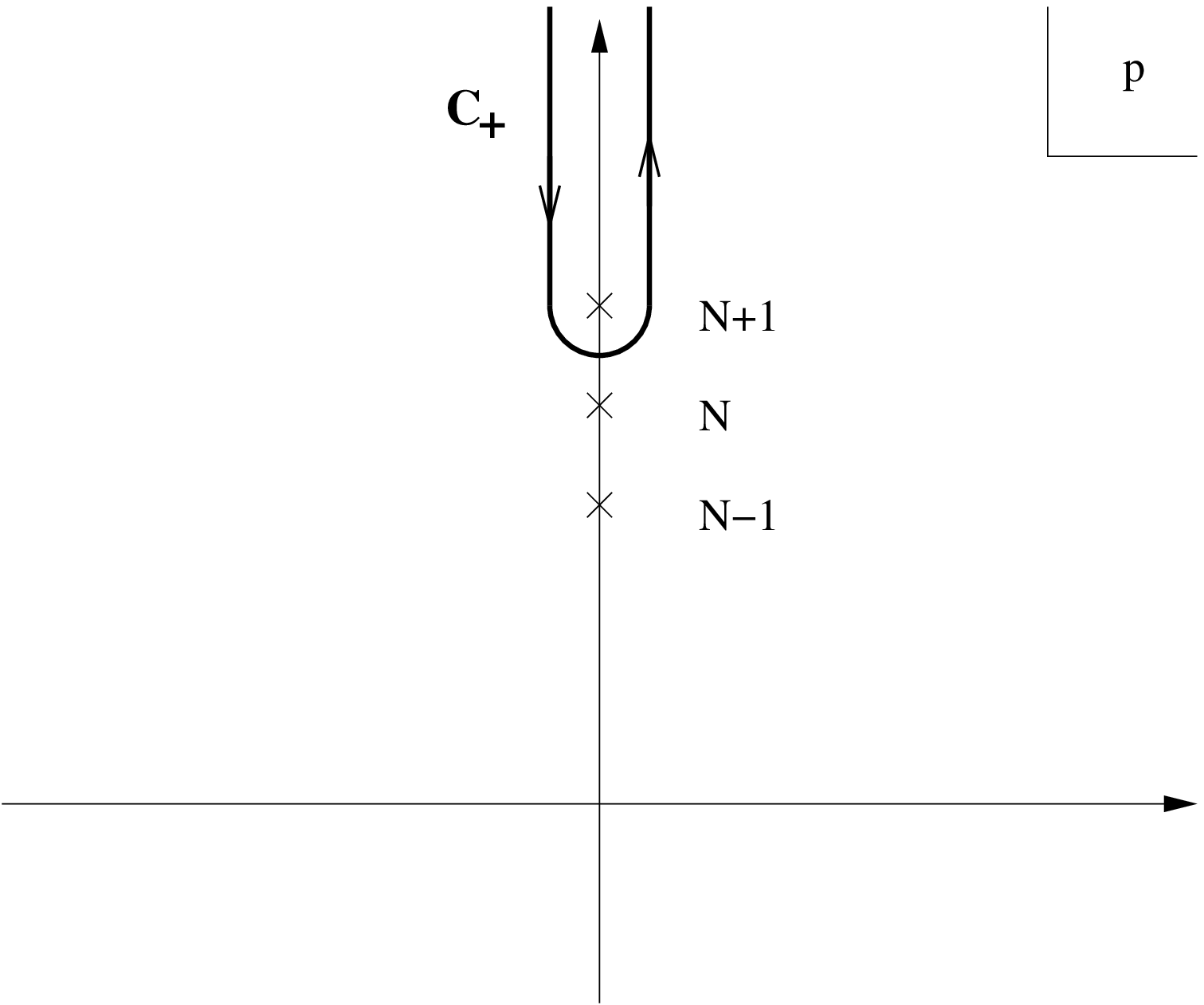}
\caption{\small The contour $C_+$} \label{contourp}
\end{minipage}
\end{figure}

Define the contour $C_-$ to be the contour coming from $-i \infty$ on the left side of the imaginary
axis, pivoting around $p=iN$ and going back down to  $-i \infty$ on the right side of the imaginary axis.
Define the contour $C_+$ to be the contour coming from $i \infty$ on the left side of the imaginary
axis, pivoting around just above $p=iN)$ and going back up to  $i \infty$ on the right side of the imaginary axis.
These are depicted in figure \ref{contourm} and figure \ref{contourp} respectively.

Now we may write,
\begin{align}
\begin{split}
 &G^{ij~\bar{T}}_{~~i'j'}\\ 
 &=C_0 \int_{C_-} \! dp \RR e^{( \! - \! N \! - \! ip)\bar{T}}
 {\Ga(-ip) \Ga(ip \! + \! N \! - \! 1) ( \! N \! + \! 1+ip) \over 2^{-2ip-1/2} ( \! N \! - ip)( \! N \! - \! 1 \! -ip)} G^{ij}_{Hi'j'} (l, \! N \! +ip) \\
 &+C_0 \int_{C_+} \! dp \RR e^{( \! - \! N \! - \! ip)\bar{T}}
   {\Ga(ip) \Ga(-ip \! + \! N \! - \! 1) ( \! N \! + \! 1-ip) \over 2^{2ip-1/2} ( \! N \! +ip)( \! N \! - \! 1 \! +ip)} G^{ij}_{Hi'j'} (l, \! N \! -ip) \\
 &\equiv I_- + I_+
\end{split}
\label{updown1}
\end{align}
Note that we have gotten rid of the hat on the propagator by multiplying it by $e^{-N \bar{T}}$.
 
The poles of the integrand of $I_+$ are given as the following.
\begin{enumerate}
    \item $p=in$ for integers $n$.
    \item $p=iN$, $p=i(N-1)$
    \item $p=-i(N-1+n)$ for non-negative integer $n$ other than $n=2$.
    \item The poles of $\sR$ (including $p=iN$).
\end{enumerate}
The non-negative integer poles come from the gamma function while the negative
integer poles come from the poles of $G^{ij}_{H~i'j'} (l,N-ip)$.
Note that these poles may `pile up.' For example, when $N$ is an integer,
the pole $p=iN$ becomes a triple pole due to the $p=in$ pole of the first line,
the $p=iN$ pole of the second line, and the $p=iN$ pole that comes from the reflection
coefficient. Note that this is written for the general case. For special values of $X_0$
the zeros coming from $\sR$ may cancel some poles mentioned above.
For reasons evident later, we mention the behavior of
the integrand of $I_+$ at $p=iN$ and $i(N-1)$.
\begin{enumerate}
    \item $p=iN$ is a triple(double) pole for integer(half-integer) $N$.
    \item $p=i(N-1)$ is a double(single) pole for integer(half-integer) $N$.
\end{enumerate}

The poles of the integrand of $I_-$ are given as the following.
\begin{enumerate}
    \item $p=in$ for integers $n$.
    \item $p=-iN$, $p=-i(N-1)$
    \item $p=i(N-1+n)$ for non-negative integer $n$ other than $n=2$.
    \item The poles of $\sR$ (including $p=iN$).
\end{enumerate}
The non-positive integer poles come from the gamma function while the positive
integer poles come from the poles of $G^{ij}_{H~i'j'} (l,N+ip)$.
The features discussed about the latter piece apply to this piece as well.
One notable feature in this case is that $p=-iN$ always turns out to
be a simple pole. To elaborate, for integer $N$, we get the contributions of the first line
and second line to get a double pole at $-iN$ while a zero at $-iN$ for $\sR$ appears to
make the pole simple. This zero in $\sR$ doesn't exist for half-integer $N$, making the pole
simple also in this case.
The behavior of the integrand of $I_-$ at $p=\pm iN$ and $\pm i(N-1)$ are as the following.
\begin{enumerate}
    \item $p=iN$ is a triple(double) pole for integer(half-integer) $N$.
    \item $p=-iN$ is always a single pole.
    \item $p=i(N-1)$ is a double(single) pole for integer(half-integer) $N$.
    \item $p=-i(N-1)$ is a double(single) pole for integer(half-integer) $N$.
\end{enumerate}

$I_+$ can be written easily as we don't have to deal with any double poles.
\be
 I_+ = \sum_{n= [N]+1 }^{\infty} A_n e^{(-N+n)\bar{T}} G^{ij}_{H~i'j'} (l, N+n)
\ee

$I_-$ has some double poles we have to think about. The simple pole contribution can be written as,
\begin{align}
\begin{split}
 I_{-,1} = &\sum_{n=1}^{[N]} A_n e^{(-N+n)\bar{T}} G^{ij}_{H~i'j'} (l, N+n) \\
         + &\sum_{n= -\infty}^{0} B_n e^{(-N+n)\bar{T}} G^{ij}_{H~i'j'} (l, N-n) \\
         + &\sum_{ia_n :~\text{poles of }\RR;~ a_n <N } C_n e^{(-N+a_n)\bar{T}} G^{ij}_{H~i'j'} (l, N-a_n) \\
         + &\delta_{N,[N]+1/2} (B_{N} G^{ij}_{H~i'j'} (l,0) + B_{(N-1)} e^{-\bar{T}} G^{ij}_{H~i'j'} (l, 1) \\
         + &B_{-(N-1)} e^{-(2N-1)\bar{T}} G^{ij}_{H~i'j'} (l,2N-1)+B_{-N} e^{-2N\bar{T}} G^{ij}_{H~i'j'} (l,2N))
\end{split}
\end{align}

For odd dimensions, we always get the double pole at $p=iN$;
\begin{align}
\begin{split}
 I_{-,2,odd} = &J_N \bar{T} G^{ij}_{H~i'j'} (l,0)
              + K_N {\p \ov \p \Delta } G^{ij}_{H~i'j'} (l,\Delta) |_{\Delta=0}
\end{split}
\end{align}
Note that,
\be
 {\p \ov \p \Delta } G^{ij}_{H~i'j'} (l,\Delta) |_{\Delta=\Delta_0}
 \sim le^{-\Delta_0 l}t^{ij}_{~~i'j'}\quad \text{for large }l
\ee

For even dimensions, we get double poles at $p=\pm i(N-1)$ and a triple pole at $p=iN$.
The double poles give rise to the terms,
\begin{align}
\begin{split}
 I_{-,2,even} = &D_{N-1} \bar{T} e^{-\bar{T}} G^{ij}_{H~i'j'} (l, 2N-1) + B^H_{N-1} e^{-\bar{T}} H^{ij}_{0~i'j'} (l, 1) \\
              +&J_{-(N-1)} \bar{T} e^{-(2N-1)\bar{T}} G^{ij}_{H~i'j'} (l, 2N-1) \\
              +&K_{-(N-1)} e^{-(2N-1)\bar{T}} {\p \ov \p \Delta } G^{ij}_{H~i'j'} (l,\Delta) |_{\Delta=2N-1}
\end{split}
\end{align}
and the triple pole gives rise to the term,
\begin{align}
\begin{split}
 I_{-,3,even} = &D_N \bar{T} G^{ij}_{H~i'j'} (l,2N) + F_N \bar{T}^2 G^{ij}_{H~i'j'} (l,2N) \\
               +&B^H_N H^{ij}_{0~i'j'} (l,0) + J^H_N \bar{T} H^{ij}_{0~i'j'} (l,0)+ K^H_N H^{ij}_{1~i'j'} (l,0)
\end{split}
\end{align}
Note that in the even dimensional case we have neglected the pieces already put into $I_{-,1}$.
The definition for the functions $H_0$ and $H_1$ are given in appendix \ref{ap:singprop},
by equations (\ref{H0}) and (\ref{H1}).

We see that in both the odd and even dimensional case, the asymptotic behavior of
the propagator is logarithmic, that is that it behaves as $\sim l t^{ij}_{~~i'j'}$.

\section{Gauge Choice} \label{s:gauge}

In the previous section, we have obtained the expression for
the transverse traceless graviton propagator.
As previously mentioned at the beginning of section \ref{s:grav1},
the transverse traceless perturbation of the graviton is
`almost' gauge invariant, that is, the transeverse tracelessness
fixes the gauge degrees of freedom except with respect to a few modes.

In section \ref{ss:gcci} we will elaborate on what we mean by
saying that there exists residual gauge freedom.
In this section we will also present a `naive' way of
getting rid of those gauge degrees of freedom.
We will present the propagator
that is gauge-fixed in this manner in \ref{ss:largel2}.

Finally, in section \ref{ss:doublepoles} we will
discuss the subtlety overlooked in the gauge-fixing method
presented in the first subsection and present
what we believe is to be the correct gauge-fixed propagator.

\subsection{Gauge Choice and Contour Integration} \label{ss:gcci}

An important issue we must address is the residual gauge degrees of
freedom we haven't gotten rid of in calculating the graviton correlator.
In other words, we have to get rid of ``degenerate modes"
of the transverse-traceless graviton.

In order to clarify what `degenerate' means,
we first decompose the graviton in our background.
We know that the (perturbation of the) graviton on a $H^{D-1}$
slice of the bubble can be (almost) uniquely decomposed as,
\be
 \delta g_{ij} = {1 \ov D-1} h \tilde{\gamma}_{ij} + 2 (\tilde{\nabla}_i \tilde{\nabla}_j
                 - {\tilde{\gamma}_{ij} \ov D-1} \widetilde{\Box}) E + 2F_{(i|j)} + h_{ij}
\label{decomp}
\ee
Where $\tilde{\gamma}_{ij}$ is the unit $H^{D-1}$ metric
(see for example, \cite{Hawking:2000ee}.)
We have used $|j$ as a shorthand for $\tilde{\nabla}_j$.
Here $h, E$ are scalars, $F_i$ is a transverse vector,
and as we know, $h_{ij}$ is a transverse traceless symmetric tensor.
We have stated that we are only interested in the
two point function of the $h_{ij}$ perturbation.

Hence the path integral we carry out concerns modes of
the transverse traceless perturbation on $H^{D-1}$.
Note that,
\be
 \Psi^h_{p'} (T+i \pi/2) r^{(pu)ij} (\HH)
\ee
would serve as an orthonormal basis of such modes,
where $r^{(pu)ij}$ are transeverse traceless eigenmodes of
\begin{align}
	\widetilde{\Box} r^{(pu)ij} = -(N^2 + 2 + p^2 ) r^{(pu)}_{i'j'}
\label{req}
\end{align}
which are normalized so that
\begin{align}
	\int d^{D-1} x \sqrt{\tilde{\gamma}} r^{(pu)ij} r^{(p'u')\dagger}_{ij} = \delta(p-p') \delta^{uu'}
\label{normalized}
\end{align}
in $H^{D-1}$. Note that $\widetilde{\Box}$ is the Laplacian
with respect to $\tilde{\gamma}_{ij}$, $\tilde{\gamma}=\det \tilde{\gamma}_{ij}$,
and as before, $u$ denotes quantum numbers other than $p$.
$\Psi^h_{p'} (X)$ are defined in (\ref{Psi}).

The problem is that there are modes that introduce an ambiguity to
the decomposition (\ref{decomp}).
Suppose there is a transverse mode $F^{(pu)i}$ such that,
$F^{(pu)(i|j)}$ is transverse, traceless and satisfies,
\begin{align}
 \widetilde{\Box} F^{(pu)(i|j)} = -(N^2 + 2 + p'^2 ) F^{(pu)(i|j)}
\end{align}
Then for this perturbation of the graviton in the angular direction,
it is ambiguous whether to put
\be
 2F^i = f(T) F^{(pu)i}
\ee
or to put,
\be
 h^{ij} = f(T) F^{(pu)(i|j)}
\ee
where $f(T)$ an arbirary function only of $T$.
The same is true if we had a scalar mode $E^{pu}$ such that,
$E_{;ij}-{\tilde{\gamma}_{ij} \ov D-1}E^{;i}_{~;i}$ is transverse traceless and
satisfies,
\begin{align}
 \widetilde{\Box} (E_{;ij}-{\tilde{\gamma}_{ij} \ov D-1}E^{;i}_{;i})
 = -(N^2 + 2 + p''^2 ) (E_{;ij}-{\tilde{\gamma}_{ij} \ov D-1}E^{;i}_{~;i})
\end{align}
For this perturbation in the angular direction,
it is ambiguous whether to put,
\be
 2E = f(T) E^{(pu)}
\ee
or to put,
\be
 h_{ij} = f(T) (E_{;ij}-{\tilde{\gamma}_{ij} \ov D-1}E^{;i}_{~;i})
\ee
This signals a `degeneracy' in the vector/scalar and tensor modes
of the graviton. By `degenerate modes' we are refering to
these modes that may be written in terms of other components
in the decomposition (\ref{decomp}).

The statement we have made in section \ref{s:grav1} that we will
only consider transverse traceless perturbations is actually a gauge
condition; that $E=0$ and $F^i=0$.
Hence such modes of $h^{ij}$ represent a residual
gauge freedom we haven't fixed yet,
as these may well be written as perturbations of the scalar/vector modes.
Therefore, in order to completely fix the gauge,
we should find them and project them out.

We will check in the appendix \ref{ap:degmodes} that the ``supercurvature modes"
$p=iN$ and $p=i(N-1)$ are degenerate with vector modes and the scalar mode
respectively.
Let's see how to project these out from the propagator.

We first start from (\ref{Greenfinal1}).
We can write this in a more convenient manner similar to (\ref{Greenprelim}),
which is,
\begin{align}
 G^{ij~\bar{T}}_{~~i'j'} (T,T', l)= \int_{C} dp (\frac{i}{2p})
  {e^{ip(2X_0-i\pi - \bar{T})} \over 1-e^{2 \pi (p-iN)}} \sR(p)  W^{ij}_{(p)i'j'} (il) 
\label{Greenprelim2}
\end{align}
In order to see how the individual tensor modes on $H^{D-1}$ contribute to
this propagator, we have to go through some steps.

We first define the maximally symmetric bitensor,
\begin{align}
	Z^{ij}_{(p)i'j'}(l) = \sum_{u} r^{(pu)ij}(\mathcal{H})^\dagger r^{(pu)}_{i'j'}(\mathcal{H}')
\label{defZ}
\end{align}
From the general prescription of obtaining maximally symmetric bitensors
which come from the sum of well defined modes in $S^d$ and $H^d$
(which is kindly laid out for the case $d=3$ in \cite{Allen:1994yb},)
we know that the relation,
\be
 Z^{ij}_{(p)i'j'} (l) ={Q'_p \ov Q_p} W^{ij}_{(p)i'j'} (il)
\ee
holds.
This is more explicitly addressed in \cite{Camporesi:1994ga},
where a multiple of $Q'_p$ is denoted as a `spectral function'.
From equation (2.107) in this paper, we see that
\be
 Q'_p = {D[p^2+(N+1)^2] \ov 2^{D-1} \pi^{N+1/2} \Ga (N+1/2) } {\Ga(ip+N-1) \Ga(-ip+N-1) \ov \Ga(ip) \Ga(-ip)}
\ee

The problems is that ${Q'_p /Q_p}$ turns out to have simple poles for
$p=i(N-1), iN$ and $p=i(N+2), i(N+3), \cdots$.
(We will only be concerned with the first two poles, as they are the ones relevant 
to the contour integral.)
We must address how to think about the pole of $Z^{ij}_{(p)i'j'}(l)$.

The poles of $Z^{ij}_{(p)i'j'}$ come from the normalization constant
of the individual modes that diverge for the given values of $p$
(see \cite{Camporesi:1994ga}.)
Since $W^{ij}_{(p)i'j'} (il)$ is obtained by multiplying an analytic function
of $p$ to get rid of the poles in the upperhalf plane, it can be written as,
\begin{align}
 W^{ij}_{(p)i'j'} (il) = \sum_{u : \text{non-zero r'}} r'^{(pu)ij\dagger}(\mathcal{H}) r'^{(pu)}_{i'j'}(\mathcal{H}')
\label{defZpr}
\end{align}
where $r'^{(pu)ij}$ aren't normalized properly.
This means that for certain values of $p$ and $u$,
$r'^{(pu)ij}$ may be zero.
This is because symmetric transverse traceless tensor modes on $H^{D-1}$
have different normalization constants for different quantum numbers.
For example, the parity even spin 2 tensor modes of have an
extra factor of $1/\sqrt{p^2+(N-1)^2}$ in their normalization constant
compared to the parity odd spin 2 tensor modes on $H^{D-1}$.
\footnote{See section 2 of \cite{Camporesi:1994ga} for more details.}
We have modified the sum over $u$ to make this point clear.
To state this more clearly,
$\{ r'^{(pu)ij} \} \subset \{ r^{(pu)ij} \}$ and
in some cases, $\{ r'^{(pu)ij} \} \neq \{ r^{(pu)ij} \}$.
We also note that,
\begin{align}
 \p_p W^{ij}_{(p)i'j'}(l) = \sum_{u : \text{non-zero }r'} 
 (\p_p r'^{(pu)ij\dagger}(\mathcal{H}) r'^{(pu)}_{i'j'}(\mathcal{H}')
 +  r'^{(pu)ij\dagger}(\mathcal{H}) \p_p r'^{(pu)}_{i'j'}(\mathcal{H}'))
\label{defZpr2}
\end{align}
We stress again that $W^{ij}_{(p)i'j'}(l)$ is well defined(regular) in the upper half plane.

Now we can write (\ref{Greenprelim2}) as,
\begin{align}
\begin{split}
 &G^{ij~\bar{T}}_{~~i'j'} (T,T', l)= \int_{-\infty}^\infty dp (\frac{i}{2p})
  {e^{ip(2X_0-i\pi - \bar{T})} \over 1-e^{2 \pi (p-iN)}} \sR(p)  W^{ij}_{(p)i'j'} (il) \\
  &-2\pi i \sum_{\substack{p_R = \{a_n \} \\ \{(N-[N-1/2]),\cdots,N \}} } \text{Res}_{p=ip_R}
  (\frac{i}{2p}) {e^{ip(2X_0-i\pi - \bar{T})} \over 1-e^{2 \pi (p-iN)}} \sR(p)  W^{ij}_{(p)i'j'} (il)
\end{split}
\end{align}
where $ia_n$ are the positive poles of the reflection coefficient.
This can be schematically written as,
\begin{align}
\begin{split}
  G^{ij~\bar{T}}_{~~i'j'} (T,T', l)= &\sum_{p : \text{real}} \Phi_p (\bar{T}) Z^{ij}_{(p)i'j'}(l) \\
                                     &+ \sum_{p = ia_n, i(N-[N-1/2]), \cdots i(N-2)} \Phi_p (\bar{T}) Z^{ij}_{(p)i'j'}(l) \\
                                     &+ \Phi_{i(N-1)} (\bar{T}) W^{ij}_{(i(N-1))i'j'}(l) \\
                                     &+ \Phi_{iN} (\bar{T}) W^{ij}_{(iN)i'j'}(l) + \Phi'_{iN} (\bar{T}) \p_p W^{ij}_{(p)i'j'}(l) |_{p=iN}
\end{split}
\label{modesum}
\end{align}
The $\Phi_{p}(\bar{T})$ denotes the $\bar{T}$ dependence of each component.
Note that for certain values of $X_0$, $in$ and $ia_n$ can coincide to give multiple poles,
but this is irrelevant to the point we wish to make now, so we will ignore such subtleties.
Note that the last line two lines come from the poles at $p=i(N-1), iN$.
From (\ref{defZ}), (\ref{defZpr}) and (\ref{defZpr2}) we see that the
expression (\ref{modesum}) shows explicitly the contribution of each hyperbolic mode to the propagator.

In appendix \ref{ap:degmodes}, it is shown that indeed the mode sum (\ref{defZpr}) for $p=i(N-1), iN$
can be written as a sum of modes coming from scalar and vector modes.
Although the scalar and vector
modes might not saturate $\{ r^{(pu)ij} \}$ (as we see in the appendix, the scalar mode derivatives only
give rise to the even tensor modes), it certainly saturates $\{ r'^{(pu)ij} \}$ as some of the
$r^{(pu)ij}$ obtain zero coefficients for the given $p$.
(This degeneracy is explicitly verified for the 4 dimensional case in \cite{Tanaka:1997kq}.)

\begin{figure}[!t]
\centering \includegraphics[width=10cm]{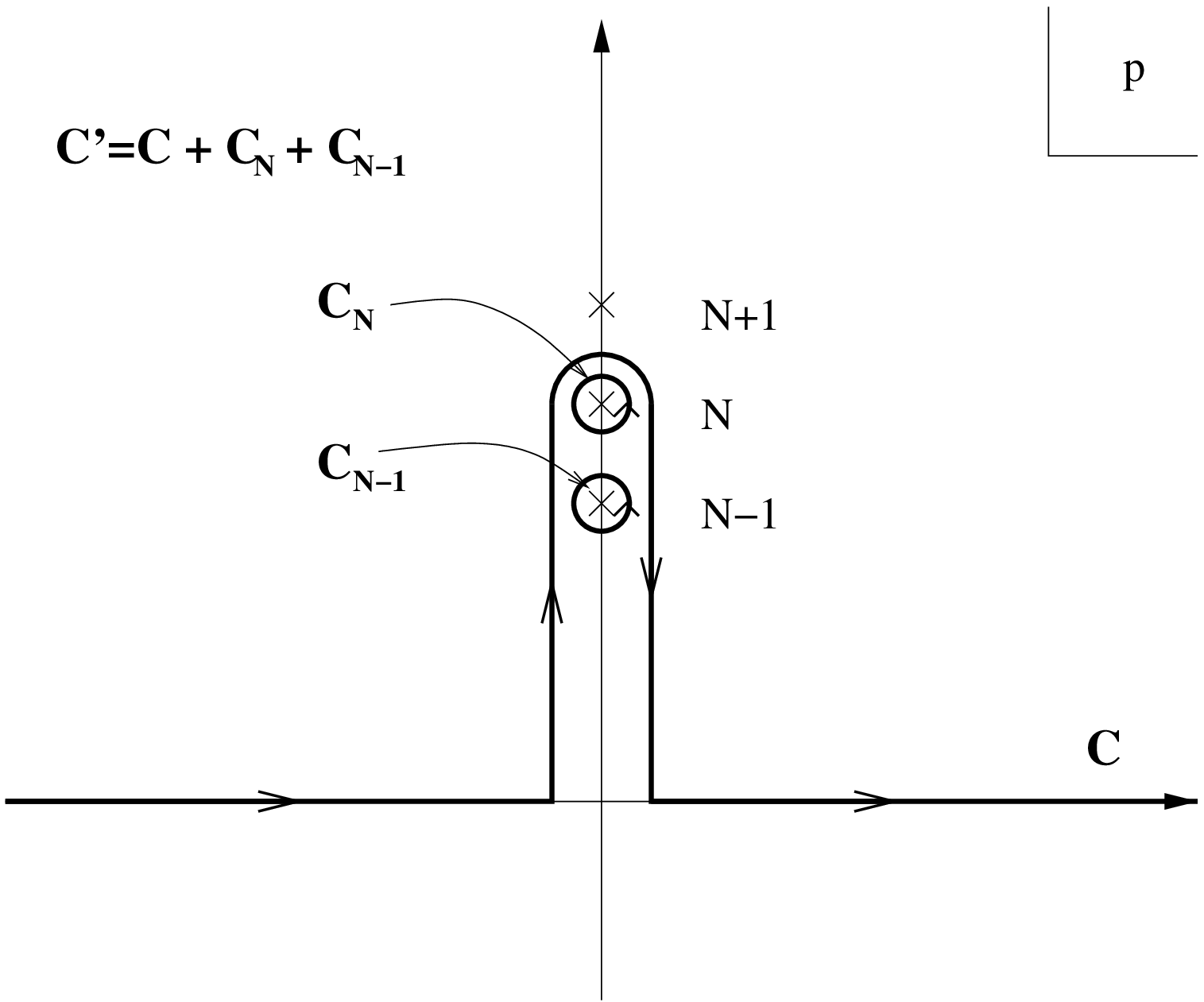}
\caption{\small The contour $C'$}\label{contourN}
\end{figure}

The naive way to project out the degenerate modes
would be to not sum over the modes of the graviton
whose `angular' modes on $H^{D-1}$ are $r'^{(pu)ij}~(p=i(N-1), iN)$ 
in the path integral in the first place.
This can be done by taking the $r'^{(pu)ij}$ components with
$p=i(N-1), iN$ in the sum (\ref{modesum}) to be zero. This just gives us,
\begin{align}
\begin{split}
  G^{ij~\bar{T}}_{~~i'j'} (T,T', l)= &\sum_{p : \text{real}} \Phi_p (\bar{T}) Z^{ij}_{(p)i'j'}(l) \\
                                     &+ \sum_{p = ia_n, i(N-[N-1/2]), \cdots i(N-2)} \Phi_p (\bar{T}) Z^{ij}_{(p)i'j'}(l) \\
\end{split}
\end{align}
which can be obtained by deforming the initial contour of integration in
(\ref{Greenfinal2}) to be $C'$ which is
$C$ with two circular contours in the counter-clockwise direction centered at $p=i(N-1)$ and $iN$ added.
(We will call these two circles $C_N$ and $C_{N-1}$ respectively.)
This is depicted in figure \ref{contourN}.
Note that if there are no poles between $i(N-1)$ and $iN$ coming from the reflection coefficient,
$C'$ can be taken to be a contour that runs along the real axis of the $p$ plane, with a `jump'
that just passes under $p=i(N-1)$.

Hence the propagator
with the redundant modes naively projected out is,
\begin{align}
\begin{split}
 G^{ij~\bar{T}}_{P~i'j'} (T,T', l) = C_0 &\int_{C'} dp 
  \RR e^{-ip \bar{T}} Y^{ij}_{(p)i'j'} (il) \\
 \times & (p^2+(N+1)^2) \Ga(ip+N-1)\Ga(-ip+N-1)
\label{Greenfinal2}
\end{split}
\end{align}

\subsection{The Large $l$ Limit (Again)} \label{ss:largel2}

Notice that projecting out the given modes do not change the arguments given
in section \ref{ss:largel} that much. Notice that equation (\ref{updown})
just gets modified by redefining the contour of integration.
That is, $I_+$ becomes $I'_+$ where we have the same integrand as $I_+$
with the different contour, $C_+' \equiv C_+ + C_N+ C_{N-1}$.
Also, $I_-$ becomes $I'_-$ which is the integral with
the same integrand as $I_-$ but with the different contour of integration,
$C_-' \equiv C_- -C_N- C_{N-1}$.

Since only the poles, $p=iN$ and $p=i(N-1)$ cross over from $I_-$ to $I_+$,
we can figure out $I'_-$ and $I'_+$ easily. First of all,
\begin{align}
\begin{split}
 &I'_{+,1} + I'_{-,1} \\
 &= \sum_{n= 1 }^{\infty} A'_n e^{(-N+n)\bar{T}} G^{ij}_{H~i'j'} (l, N+n) \\
 &+ \sum_{n= -\infty}^{0} B'_n e^{(-N+n)\bar{T}} G^{ij}_{H~i'j'} (l, N-n) \\
 &+ \sum_{ia_n :~\text{poles of }\RR;~ a_n <N } C_n e^{(-N+a_n)\bar{T}} G^{ij}_{H~i'j'} (l, N-a_n) \\
 &+ \delta_{N,[N]+1/2} (A'_N G^{ij}_{H~i'j'} (l, 2N) + A'_{N-1} e^{-\bar{T}} G^{ij}_{H~i'j'} (l, 2N-1) \\
 & +B'_{-(N-1)} e^{-(2N-1)\bar{T}} G^{ij}_{H~i'j'} (l,2N-1)+B'_{-N} e^{-2N\bar{T}} G^{ij}_{H~i'j'} (l,2N))
\label{fin1}
\end{split}
\end{align}
We note that $A'_n=A_n$ and $B'_n=B_n$ for $n \neq N, \pm(N-1)$.

For odd dimensions, we get the double pole at $p=iN$ in $I'_+$;
\begin{align}
\begin{split}
 I'_{+,2,odd} = &D'_N \bar{T} G^{ij}_{H~i'j'} (l,2N)
               + E'_N {\p \ov \p \Delta } G^{ij}_{H~i'j'} (l,\Delta) |_{\Delta=2N}
\label{fin2odd}
\end{split}
\end{align}

For even dimensions, we get double poles at $p=\pm i(N-1)$ and a triple pole at $p=iN$.
The double poles give rise to the terms,
\begin{align}
 I_{+,2,even} = &D'_{N-1} \bar{T} e^{-\bar{T}} G^{ij}_{H~i'j'} (l, 2N-1)
                + E'_{N-1} e^{-{\bar{T}}}{\p \ov \p \Delta } G^{ij}_{H~i'j'} (l,\Delta) |_{\Delta=2N-1}\\
 I_{-,2,even} = &J'_{-(N-1)} \bar{T} e^{-(2N-1)\bar{T}} G^{ij}_{H~i'j'} (l, 2N-1) \\
              + &K'_{-(N-1)} e^{-(2N-1)\bar{T}} {\p \ov \p \Delta } G^{ij}_{H~i'j'} (l,\Delta) |_{\Delta=2N-1}
\label{fin2even}
\end{align}
and the triple pole gives rise to the term,
\begin{align}
\begin{split}
 I_{-,3,even} = &D'_N \bar{T} G^{ij}_{H~i'j'} (l,2N) + F'_N \bar{T}^2 G^{ij}_{H~i'j'} (l,2N) \\
               +&(E'_N {\p \ov \p \Delta } G^{ij}_{H~i'j'} (l,\Delta)
                 + G'_N \bar{T} {\p \ov \p \Delta } G^{ij}_{H~i'j'} (l,\Delta) \\
                &+H'_N {\p^2 \ov \p \Delta^2 } G^{ij}_{H~i'j'} (l,\Delta) )|_{\Delta=2N}
\label{fin3even}
\end{split}
\end{align}

Note that both in the even and odd dimensional case, we have tamed the logarithmic
scaling behavior. Now the asymptotic behavior goes in general like, $\sim e^{-(N-a_n)l}$
where $ia_n$ is the pole of $\sR$ with maximum $\Re a_n$, or if there aren't any poles
of $\sR$ in the upper half plane, it would be like $\sim e^{-Nl}$.

\subsection{Treatment of Double Poles} \label{ss:doublepoles}

It seems that we have obtained a well tamed propagator in the previous section,
but there is a subtlety we have overlooked.
This comes from the fact that we have ignored the contribution of the
double pole in (\ref{modesum}). We have done this by assuming that
we can ignore the contribution of gauge dependent modes in the propagator,
since these modes can show up by a coordinate transformation.

But it is not clear that this is the thing to do when our
propagator is not `diagonal.'
The problem is that the statement that
`we ignore the gauge dependent modes' just restricts
the form of ``$h^{ij}$", so schematically, if we denote the
modes that should be projected out to be, $\ket{p}$,
the gauge condition is just,
\be
 \langle p \ket{h}=0
\label{gcon}
\ee
for the states $\ket{h}$.

Now if our unprojected propagator is diagonal, i.e. of the form,
\be
 G = \sum_m M_m \ket{m}\bra{m}
\ee
to begin with, the gauge condition can be translated into,
\be
 G = \sum_{m \neq p} M_m \ket{m}\bra{m}
\ee
since
\be
 G \ket{h} = \sum_m M_m \ket{m} \langle m \ket{h} = \sum_{m \neq p} M_m \ket{m} \langle m \ket{h}
\ee
for $\ket{h}$ satisfying the gauge condition (\ref{gcon}) anyways.

But if the propagator, as in our case, had the form,
\be
 G = \sum_{m \neq p} M_m \ket{m}\bra{m} + M_0 \ket{0}\bra{0}+ M_1 ( \ket{0'}\bra{0}+\ket{0}\bra{0'})
\label{G}
\ee
where $\ket{0}$ is a mode we projected out, still,
\be
 G\sum_{m \neq p} a_m \ket{m} = \sum_{m \neq p} M_m a_m \ket{m} + \sum_{m \neq p} M_1 a_m \langle 0' \ket{m} \ket{0}
\ee
so we would have to keep the latter term with $M_1$,
since it has a physical effect on the gauge fixed states.

Suppose the propagator $G$ of the form (\ref{G})
could be written in the particular form,
\be
 G = \sum_{m \neq p} M_m \ket{m}\bra{m} + M_0 \ket{0}\bra{0}+ M_1 \lim_{\epsilon \ra 0}
 ({1 \ov \epsilon}\ket{0}\bra{0} - {1 \ov \epsilon}\ket{\epsilon}\bra{\epsilon})
\label{gmod}
\ee
where,
\be
 \lim_{\epsilon \ra 0} \ket{\epsilon} = \ket{0}
\ee

In this case,
\be
 \ket{0'} = -\lim_{\epsilon \ra 0} {\ket{\epsilon} - \ket{0} \ov \epsilon}
\ee
Actually, this is exactly what happens to our propagator as we
deform the reflection coefficient of the thin wall.
Even if we project out $\ket{0}$, we would still have to keep $\ket{\epsilon}$,
but the last term in (\ref{gmod}) is not well defined, so we regulate it by subtracting
${1 \ov \epsilon}\ket{0}\bra{0}$ and obtain the $M_1$ term of (\ref{G}).

So the reason that this piece shows a physical effect becomes
clearer in our case.
Although the bubble wall bound state mode becomes degenerate with a
gauge mode, we cannot treat it as if it did not exist in the first place.

Retaining the double pole contribution,
we should write the propagator as,
\begin{align}
\begin{split}
 &G^{ij~\bar{T}}_{P~i'j'} (T,T', l) = \\
 &\ C_0 \int_{C'} dp \RR e^{-ip \bar{T}} Y^{ij}_{(p)i'j'} (il) (p^2+( \! N \! + \! 1)^2) \Ga(ip \! + \! N \! - \! 1)\Ga(-ip \! + \! N \! - \! 1) \\
 &\ + K_N \p_p W^{ij}_{(p)i'j'}(l)|_{p=iN}
\label{Greenfinal3}
\end{split}
\end{align}

It is easily verifiable that for large $l$ (and small $z$ in poincare coordinates,)
\be
 \p_p W^{ij}_{(p)i'j'}(l)|_{p=iN} \sim l t^{ij}_{~~i'j'} \sim \ln(|x-x'|/z) t^{ij}_{~~i'j'}
\ee


The boundary curvature two point function arising
from this piece is non-zero.
We will do this calculation in section \ref{ss:log}.

\section{Summary} \label{s:obs}

We have seen in the previous section that we can organize the
propagators in the large $l$ limit as a sum of
well defined transverse traceless propagators in $H^{D-1}$
coming from single poles in the momentum integral,
their normalizable derivatives,
and a non-normalizable logarithmic piece
coming from the double pole.
Put in this form, hopefully, it should be easier
to think about what a holographic theory on the $S^{D-1}$ boundary
of the ``bubble", if exists, would look like.

In this section, we will takes steps to further carry out this effort.
We will first summarize the graviton two point
function; we will sort out the terms in a well organized way.
We will also point out some important features that
may have implications about the boundary theory.

\subsection{Summary of Results} \label{ss:sumgrav}

Let's once and for all write down the terms that show up
in our `holographic expansion' of the
gauge invariant transverse traceless graviton two point function
in a $D$ dimensional CDL instanton background.
The full two point function can be written as,
\be
G^{ij}_{~~i'j'} (T,T', l) = G^{ij~\delta T}_{~~i'j'} (T,T', l) + G^{ij~\bar{T}}_{P~i'j'} (T,T', l)
\ee
where $G^{ij~\delta T}_{~~i'j'} (T,T', l)$ is the graviton two point function
in flat FRW space with no bubble nucleation.
We are only intrested in the piece,
$G^{ij~\bar{T}}_{P~i'j'} (T,T', l)$
that arises due to the existence of the CDL instanton.

This can be written as,
\begin{align}
 G^{ij~\bar{T}}_{P~i'j'} (T,T', l) &= G^{single} + G^{double}
\end{align}

$G^{single}$ has the same form for both odd and even dimensions.
The form for $G^{double}$ differs according to the parity of
the dimension.

$G^{single}$ can be written as the following.
\begin{align}
\begin{split}
 G^{single}
 =&\sum_{\substack{n=0 \\ n \neq N, N-1}}^{\infty} A_n e^{-(N+n)\bar{T}} G^{ij}_{H~i'j'}(l,N+n) \\
 +&\sum_{\substack{n=0 \\ n \neq N, N-1}}^{\infty} B_n e^{-2N\bar{T}}e^{(N+n)\bar{T}} G^{ij}_{H~i'j'}(l,N+n) \\
 +&\sum_{a_n \neq N} C_n e^{-(N-a_n)\bar{T}} G^{ij}_{H~i'j'}(l,N-a_n) \\
 +&D_N \partial_p W^{ij}_{(p)i'j'}(l)|_{p=iN}
\end{split}
\end{align}
where $ia_n$ are the poles of $\sR(p)$ given by (\ref{sR}).
We once more recall that we have defined $N\equiv (D-2)/2$.

$G^{ij}_{H~i'j'}(l,\Delta)$ are propagators on $H^{D-1}$ defined by
(\ref{massprop}). For $\Delta > N$, these are proportional to
massive spin 2 propagators with $m^2 = \Delta (\Delta -2N)$.

If we write the coordinates of the points in
the hyperbolic slices using Poincar\'e coordinates,
(so the coordinates of the two point would be $(T,\vec{x},z)$ and $(T',\vec{x}',z')$,)
the propagators showing up in the above sum behave as the following as $z, z' \to 0$;
\begin{align}
\begin{split}
 (z^{N+n} z'^{N+n}) & \left[ {e^{-(N+n)T} e^{-(N+n)T'}\ov (r^2)^{(N+n)}} t^{ij}_{~~i'j'} \right] \\
 (z^{N+n} z'^{N+n})  e^{-2NT}e^{-2NT'} & \left[ {e^{(N+n)T} e^{(N+n)T'} \ov (r^2)^{(N+n)}} t^{ij}_{~~i'j'}\right] \\
 (z^{N-a_n} z'^{N-a_n}) & \left[ {e^{-(N-a_n)T}e^{-(N-a_n)T'} \ov (r^2)^{(N-a_n)}} t^{ij}_{~~i'j'} \right] \\
 & \left[ \ln(r /z) t^{ij}_{~~i'j'} \right]
\end{split}
\end{align}
where we have defined, $r\equiv |\vec{x}-\vec{x}'|$.

In even dimensions,
$G^{double}$ can be written as the following.
\begin{align}
\label{eobstr1}
\begin{split}
 G&^{double}\\
 =&E_{N-1} e^{-(2N-1)\bar{T}} [G^{ij}_{H~i'j'}(l,2N\!-\!1) (a'_0+a'_1 \bar{T})+ 
                     b'_0\partial_{\Delta} G^{ij}_{H~i'j'}(l,2N\!-\!1)] \\
 +&E_N e^{-2N\bar{T}} G^{ij}_{H~i'j'}(l,2N) \\
 +&F_{N-1} e^{-2N\bar{T}}e^{(2N-1)\bar{T}} [G^{ij}_{H~i'j'}(l,2N\!-\!1) (a_0+a_1 \bar{T})+
                                  b_0 \partial_{\Delta} G^{ij}_{H~i'j'}(l,2N\!-\!1)] \\
 +&F_{N} e^{-2N\bar{T}}e^{2N\bar{T}} [G^{ij}_{H~i'j'}(l,2N) (c_0+c_1 \bar{T}+c_2 \bar{T}^2) \\
  &\qquad \qquad \quad
  + \partial_{\Delta} G^{ij}_{H~i'j'}(l,2N) (d_0 +d_1 \bar{T})+ e_0 \partial^2_{\Delta} G^{ij}_{H~i'j'}(l,2N)]
\end{split}
\end{align}
Where we have defined,
\be
 \partial^n_{\Delta} G^{ij}_{H~i'j'}(l,\Delta') \equiv \partial^n_{\Delta} G^{ij}_{H~i'j'}(l,\Delta)|_{\Delta=\Delta'}
\ee

Taking each term to the boundary we get,
\begin{align}
\label{eobstr2}
\begin{split}
 (z^{(2N-1)} z'^{(2N-1)}) & \left[ {e^{-(2N-1)T} e^{-(2N-1)T'} \ov (r^2)^{2N-1} } t^{ij}_{~~i'j'}\right] \\
 e^{-2NT} e^{-2NT'} & \OO \left( { z^{2N+2} z'^{2N+2} \ov (r^2)^{2N+2} } \right) \\
 (z^{(2N-1)} z'^{(2N-1)}) e^{-2NT}e^{-2NT'} &
 \left[ {e^{(2N-1)T} e^{(2N-1)T'} \ov (r^2)^{2N-1}} t^{ij}_{~~i'j'} \right] \\
 (z^{2N} z'^{2N}) e^{-2NT}e^{-2NT'} &
 \left[ { e^{2N T}e^{2N T'} \ov (r^2)^{2N}} t^{ij}_{~~i'j'} \right] [(a_0+a_1 \bar{T}) +b_0 \ln ({r \ov z}) ] 
\end{split}
\end{align}

In odd dimensions,
$G^{double}$ can be written as the following.
\begin{align}
\label{oobstr1}
\begin{split}
 G&^{double} \\
 =&E_{N-1} e^{-(2N-1)\bar{T}} G^{ij}_{H~i'j'}(l,2N-1) \\
 +&E_N e^{-2N\bar{T}} G^{ij}_{H~i'j'}(l,2N) \\
 +&F_{N-1} e^{-2N\bar{T}}e^{(2N-1)\bar{T}} G^{ij}_{H~i'j'}(l,2N-1) \\
 +&F_N e^{-2N\bar{T}}e^{2N\bar{T}} [G^{ij}_{H~i'j'}(l,2N) (f_0+f_1 \bar{T})+ g_0 \partial_{\Delta} G^{ij}_{H~i'j'}(l,\Delta)|_{\Delta=2N}]
\end{split}
\end{align}

Taking each term to the boundary, we obtain,
\begin{align}
\label{oobstr2}
\begin{split}
 e^{-(2N-1)T} e^{-(2N-1)T'} & \OO \left( { z^{2N+1} z'^{2N+1} \ov (r^2)^{2N+1} } \right) \\
 e^{-2NT} e^{-2NT'} & \OO \left( { z^{2N+2} z'^{2N+2} \ov (r^2)^{2N+2} } \right) \\
 (e^{-2NT}e^{-2NT'}) e^{(2N-1)T} e^{(2N-1)T'} & \OO \left( { z^{2N+1} z'^{2N+1} \ov (r^2)^{2N+1} } \right)\\
 (z^{2N} z'^{2N}) e^{-2NT}e^{-2NT'} & \left[ {e^{2NT}e^{2NT'} \ov (r^2)^{2N}} t^{ij}_{~~i'j'} \right]
\end{split}
\end{align}

\subsection{The Logarithmic Piece} \label{ss:log}

We first focus on the piece,
\be
 \p_p W^{ij}_{(p)i'j'}(l)|_{p=iN} \sim l t^{ij}_{~~i'j'}
\label{dim0}
\ee

The natural thing to do with this is to
calculate the `curvature two point function' coming
from this piece.
To explain a bit more,
if we assume that this piece corresponds to a two point
function $<h^{ij}h_{ij}>$ of some transverse traceless operator on the boundary,
we would like to see what the gauge invariant two point function,
$\nabla_i \nabla_j \nabla^{i'} \nabla^{j'} <h^{ij}h_{i'j'}>$ is.

Let's first try to find the relevant components of this by explicitly
writing this in Poincar\'e coordinates on the hyperboloid. We do so because it is convenient
to see the behavior at the boundary in these coordinates. To write it once more,
the Poincar\'e coordinate on the hyperboloid is,
\begin{align}
  ds^2 = \frac{1}{z^2} (dz^2 + dx_1 ^2 + \cdots + dx_{D-2}^2)
\end{align}

The boundary is at $z=0$. Let's consider two points $(x_1, \cdots, x_{D-2},z)$, $(-x_1, \cdots ,-x_{D-2},z)$
and look at the $z \rightarrow 0$ limit. In the Poincar\'e coordinates, the geodesic that connects two points
on the boundary is a half circle. So considering the given two points, the unit tangent vector $n_i$ and $n_i'$
at $(x_1, \cdots, x_{D-2},z)$ and $(-x_1, \cdots ,-x_{D-2},z)$ respectively is,
\begin{align}
  \begin{pmatrix} n_z \\ n_{x_i} \end{pmatrix}
  = \begin{pmatrix} -r'/zr \\ x_i /rr' \end{pmatrix},~~ 
  \begin{pmatrix} n_{z'} \\ n_{x_i'} \end{pmatrix}
  = \begin{pmatrix} -r'/zr \\ -x_i /rr' \end{pmatrix} 
\end{align}
where we define,
\begin{align}
  r' = \sqrt{x_1^2 + x_2^2 + \cdots x_{D-2}^2}, ~~~ r = \sqrt{z^2 + s'^2}
\end{align}
for convenience.

The parallel transport operator in our case is just a rotation matrix, which is,
\begin{align}
\begin{split}
  \begin{pmatrix}
    g_z^{~z'} & g_z^{~x_j'}  \\
    g_{x_i}^{~z'} & g_{x_i}^{~x_j'}
  \end{pmatrix}
  = \frac{1}{r^2}
  \begin{pmatrix}
    -r^2 + 2z^2 & -2zx_1  & \cdots & -2zx_{D-2} \\
    2zx_1 & r^2-2x_1^2  & -\cdots & -2x_1x_{D-2} \\
    \vdots & \vdots & \ddots & \vdots \\
    2zx_{D-2} & -2x_{D-2} x_1 & \cdots & r^2 - 2 x_{D-2}^2 \\
  \end{pmatrix}
\end{split}
\end{align}


Also, the geodesic distance between the two points may be calculated as,
\begin{align}
  l = 2 \ln \frac{r+r'}{z}
\end{align}

Calculating $t^{ij}_{~~i'j'}$ from this, we find that if the correlator has at least one $z$ index, it is
of order $O(z)$. Hence, in the limit $z \rightarrow 0$, we find that the only surviving components of $t^{ij}_{~~i'j'}$
are those with all the indicies are in the $D-2$ plane on the boundary,
that is,
\be
 t^{ij}_{~~i'j'} \sim \OO(z) \qquad \text{for} ~ z \rightarrow 0 ~~ \text{unless}~i,j,i',j' \neq z
\label{tenssurv}
\ee
Hence when we take the correlator to the boundary,
the only surviving tensors components (\ref{dim0}) are those whose indices are all along the boundary directions.
Also, in this limit, $l=2 \ln (2r'/z) $

We can actually write the form of $t^{ij}_{~~i'j'}$ on the boundary plane
from direct calculation which yields,
\begin{align}
\begin{split}
  t^{ij}_{~~i'j'} ((x_1, \cdots, x_{D-2}),& (-x_1, \cdots , -x_{D-2})) \\
   =  \delta_{ij} \delta_{i'j'}
  &- N \big(\delta_{ii'} - \frac{2x_ix_{i'}}{r'^2} \big) \big( \delta_{jj'}-\frac{2x_jx_{j'}}{r'^2} \big) \\
  &- N \big(\delta_{ij'} - \frac{2x_ix_{j'}}{r'^2} \big) \big( \delta_{ji'}-\frac{2x_jx_{i'}}{r'^2} \big) + O(z)
\end{split}
\end{align}
when $i,j,i',j'$ are all along the direction of the boundary.
Using translational invariance in the boundary space we obtain,
\begin{align}
\label{deften}
\begin{split}
  &t^{ij}_{~~i'j'} ((x_1, \cdots, x_{D-2}), (y_1, \cdots , y_{D-2})) \\
  &=  \delta_{ij} \delta_{i'j'} \\
  &- N \big(\delta_{ii'} - \frac{2(x_i-y_i)(x_{i'}-y_{i'})}{R'^2} \big) \big( \delta_{jj'}-\frac{2(x_j-y_j)(x_{j'}-y_{j'})}{R'^2} \big) \\
  &- N \big(\delta_{ij'} - \frac{2(x_i-y_i)(x_{j'}-y_{j'})}{R'^2} \big) \big( \delta_{ji'}-\frac{2(x_j-y_j)(x_{i'}-y_{i'})}{R'^2} \big)
\end{split}
\end{align}
by replacing $2x_i$ by $x_i - y_i$.

Note that we have newly defined,
\begin{align}
  R'^2 = (x_1-y_1)^2 + \cdots + (x_{D-2}-y_{D-2})^2
\end{align}
which satisfies $l= 2 \ln (R'/z)$.

Let's attempt to calculate a gauge invariant quantity, the $D-2$ dimensional scalar curvature of the graviton
fluctuation. Since a traceless perturbation $h_{ij}$ of the curvature in a flat background yields,
\begin{equation}
 C \propto  \partial_i \partial_j h^{ij}
\end{equation}
we get,
\begin{align}
 <C(x ) C(y ) >  = \partial_i \partial_j \partial^{i'} \partial^{j'} c_0 l t^{ij}_{~~i'j'}
 = {c_1 \over R'^4} + {c_2 \ln (R'/z) \over R'^4 } 
\end{align}
where,
\begin{align}
c_1 &= -16(2N-1)(3N-2)(4N^2 -7N+1) c_0 \\
c_2 &= -64 N(N-1)(N-2)(2N-1)^2 c_0
\end{align}

It's worth noting that the $\ln (R'/z) / R'^4$ term vanishes only for $D=3,4,6$,
and that for $D=3$, the curvature vanishes altogether. (We've ignored the $D=2$
case since this calculation doesn't make sense if coordinates are not
defined at all in the first place.)

\subsection{Existence of a Stress-Energy Tensor} \label{ss:set}

We notice from the expression given in section \ref{ss:sumgrav}
(namely equations (\ref{eobstr1}) and (\ref{oobstr1}),) that we have a dimension $2N=(D-2)$
transverse traceless tensor propagator in piece in $H^{D-1}$.
That is, we have the pieces which in Poincar\'e coordinates,
ignoring the $T$ dependence, behaves like
\be
 \p_{\Delta} G^{ij}_{H~i'j'}(l,2N) \sim z^{2N} z'^{2N} (x-x')^{-4N} t^{ij}_{~~i'j'}
\ee
as $z,z' \rightarrow 0$.

Writing this for two points at equal $z$ we get,
\be
 \p_{\Delta} G^{ij}_{H~i'j'}(l,2N) \sim {z^{4N} \over R'^{4N}} t^{ij}_{~~i'j'}
\ee
By direct calculation, it is verified that this piece is
transverse-traceless on the $(D-2)$ dimensional boundary,
namely that,
\begin{align}
 {1 \over R'^{4N}} t^{i}_{~ii'j'} &=0 \\
 \partial_i \left( {1 \over R'^{4N}} t^{ij}_{~~i'j'} \right) &=0
\end{align}

Actually we see that this conincides
with the expression for the two point function
of the stress energy tensor of a CFT (namely equation (2.18))
given in \cite{Erdmenger:1996yc}.

We do not want to hastily imply that the piece that shows up in our
expansion is a stress engergy tensor on the boundary,
but if we assume some kind of holographic correspondence
there seems to exist a dimension $(D-2)$ transverse traceless tensor
on a $(D-2)$ dimensional boundary theory.

Note that from looking at equation (\ref{eobstr2})
there seems to be some kind of obstruction of this
term that comes from the pole $p=iN$.
We will try to address this issue in the final section.

\subsection{Odd and Even Dimensions} \label{ss:oe}

Since we know that gravity behaves very differently in even and odd dimensions,
we would expect the behavior of the propagator to be drastically different for the two cases,
which indeed it is. The most dramatic difference would be that
the number of poles of the reflection coefficient for even dimensions is finite
(as the reflection coefficient becomes a rational function with respect to $p$,)
whereas in odd dimensions it is infinite.
Hence if we want to think about some holographic correspondence,
an infinite number of operators with different dimensions seems to come at play for
odd dimensions whereas for even dimensions number seems finite.

Also in odd dimensions some values of $X_0$ seem to give rise
to an infinite number of complex poles for $\sR$.
This happens at a sharp point, namely at $X_0=0$.
If there is indeed some kind of holographic dual theory that lives at
the boundary $S^{D-2}$ that is dual to the CDL gravity theory,
this suggests that there might be some phase transition or duality
in that theory for odd dimensions, whereas for even dimensions,
where all the poles stay on the imaginary axis for all values of $X_0$,
nothing of the sort seems to happen.


\section{The Scalar Propagator} \label{s:scalar}

We will follow the exact steps taken as we have with the graviton
propagator in obtaining the propagator for
an arbitrary minimally coupled scalar $\psi$
in the given background.

\subsection{The Equation of Motion}

We first consider when $\psi$ is massless.
We first define
\be
 \chi = a^N(X) \psi
\ee
Reusing the notations we have used for the graviton,
the relevant part of the action turns out to be,
\be
 S = \frac{1}{2} \int dX d\Omega_{D-1} \sqrt{\tilde{g}} \chi [-\p^2_X +U(X) - \widetilde{\Box}] \chi
\ee

Hence by defining,
\be
 \hat{G}(X_1,X_2,\Om_1,\Om_2) = a^N(X_1 ) a^N(X_2) <\chi(X_1,\Om_1) \chi (X_2,\Om_2)>
\ee
we get,
\be
 [-\p_{X_1}^2+U(X_1)- \widetilde{\Box}_1] \hat{G}(X_1,X_2,\Om_1,\Om_2)
 = \frac{1}{\sqrt{\tilde{g}}} \delta (X_1 -X_2) \delta(\Om_1,\Om_2)
\label{Seq}
\ee

\subsection{Decomposition}

Due to the $O(D-1)$ symmetry,
the Green's function $G$ can only be
a function of $X,X'$ and the geodesic distance $\mu(\Omega_1 , \Omega_2 )$
between the two points on the $(D-1)$ sphere.
Hence, we may write the solution for the
equation (\ref{Seq}) simply as,
\begin{align}
 \hat{G} (X, X', \mu) = \sum^{+i \infty}_{ p = iN } G^s_p (X, X' ) W_{(p)}(\mu)
\label{Shateq}
\end{align}
for $G^s_p(X,X')$ and $W_{(p)}(\mu)$ which we will define below.

We define $G^s_p$ to satisfy equation (\ref{Gpeq}).
The reason we didn't just put $G^s_p$ equal to $G_p$ defined in (\ref{Gp})
is because $G_p$ obtained as (\ref{Gpfinal}) for $p=iN$ is singular
due to the pole of $\sR$ at $p=iN$.
$G_p$ has a simple pole at $p=iN$ and the residue $R(X,X')$
of this pole satisfies the equation,
\be
 [-\partial_X^2 +U(X)] R(X,X') =0
\ee
This is because $R(X,X')$ is normal at $X=X'$.
(For example, when $X,X'<X_0$ it is an exponential of $X+X'$ so
it behaves normally.)
Hence we may define
\begin{align}
 G^s_{iN} (X,X')\equiv \lim_{p \to iN} \left( G_p (X,X') -{\text{Res}_{p=iN} G_p (X,X') \ov p-iN} \right)
\end{align}
and $G^s_{iN}$ will still satisfy equation (\ref{Gpeq}) for $p=iN$.
For $p \neq iN$, we may set $G^s_p$ safely equal to $G_p$.
Hence for $X,X'<X_0$ we get,
\begin{align}
\begin{split}
 G^s_p (X,X') &\equiv \frac{i}{2p} (e^{ip \delta X} + \RR (p) e^{-ip \bar{X}}) \\
 G^s_{iN} (X,X') &\equiv \frac{i}{2p} (e^{ip \delta X} + (\RR (p)-{\text{Res}_{p=iN} \RR (p) \ov p-iN}) e^{-ip \bar{X}})
\end{split}
\label{GiN}
\end{align}

$W_{(p)}(\mu)$ is a scalar function only dependent upon $\mu(\Omega_1 , \Omega_2 )$.
$W_{(p)}(\mu)$ is defined by,
\begin{align}
	W_{(p)}(\mu) = \sum_{u} q^{(pu)}(\Omega) q^{(pu)}(\Omega')^*
\label{WSsum}
\end{align}
where $q^{(pu)}$ are transeverse traceless eigenmodes of
\begin{align}
	\widetilde{\Box} q^{(pu)} = (N^2 + p^2 ) q^{(pu)}
\end{align}
which are normalized so that
\begin{align}
	\int d^{D-1} x \sqrt{\tilde{g}} q^{(pu)} q^{(p'u')*} = \delta^{pp'} \delta^{uu'}
\end{align}
Note that we denoted all the quantum numbers other than $p$ needed to specify the mode $q$ as $u$.
$W_{(p)}(\mu)$ satisfies,
\begin{align}
	\widetilde{\Box} W_{(p)}(\mu) = (N^2 + p^2 ) W_{(p)}(\mu)
\label{Wseq1}
\end{align}

On $S^{D-1}$, we get eigenmodes for the $p$ values, $p=iN, i(N+1), \dots $, so by completeness of the basis,
\begin{align}
	\sum^{+i \infty}_{ p = iN } W_{(p)}(\mu(\Omega , \Omega' )) = \delta (\Omega , \Omega' ) 
	/\sqrt{\tilde{g}}
\label{Wseq2}
\end{align}

From equations (\ref{Wseq1}), (\ref{Wseq2}), and (\ref{Gpeq}), we see that indeed (\ref{Shateq}) solves (\ref{Seq}).

\subsection{$W_{(p)}(\mu)$}

The equation for $W_{(p)}(\mu)$ can be written out as in \cite{Allen:1985wd}, which is,
\be
  W_{(p)}''(\mu) + (D-2) \cot \mu G' (\mu) -(N^2 +p^2) G (\mu) =0
\ee
This can be solved to be,
\be
 W_{(p)} (\mu) = K_p F(N+ip,N-ip;N+{1\ov2};1-z) \quad \text{for}\quad z= \cos^2 {\mu \ov 2}
\label{WSsemfin}
\ee
where from (\ref{WSsum}) we see that $W_p$ to be non-singular at $\mu =0$.
$K_p$ can be calculated from (\ref{WSsemfin}) and (\ref{WSsum})
\begin{align}
\begin{split}
  K_p {2\pi^{D/2} \ov \Ga(D/2)} &= \int d^{D-1} \Om \sqrt{\tilde{\gamma}} W_{(p)} (\Om, \Om) \\
  &= -{2ip(p^2+(N-1)^2) \Ga(-ip+N-1) \ov (D-2)! \Ga(-ip-N+2)}
\end{split}
\end{align}
by the degeneracy of the $p$ mode \cite{Camporesi:1994ga}.
Hence we obtain,
\begin{align}
\begin{split}
 W_{(p)} (\mu) = &[-{i\Ga(D/2)\ov \pi^{D/2}}] {p(p^2+(N-1)^2) \Ga(-ip+N-1) \ov (D-2)! \Ga(-ip-N+2)} \\
 &\times F(N+ip,N-ip;N+{1\ov2};1-z)
\label{WSfin}
\end{split}
\end{align}

\subsection{Massive Scalar Propagators in $H^{(D-1)}$}

The equation for the propagator for a massive scalar
in $H^{D-1}$ with curvature radius $R^2=-1$ is,
\be
 (-\widetilde{\Box}_1-m^2) G_H (l(\Omega_1, \Omega_2),m^2) =
 {1 \over \sqrt{\tilde{\gamma}}} \delta (\Omega_1, \Omega_2)
\ee

This is solved in \cite{Allen:1985wd} to be,
\begin{align}
\begin{split}
  G_M& (l,m^2) = \\
 &[{\Ga(N-ip) \Ga(1/2-ip) \ov \Ga(1-2ip) \pi^{(D-1)/2} 2^{(D-1)}} ]
                ({1 \ov z})^{N-ip} F(N-ip,1/2-ip;1-2ip;{1 \ov z})
\label{smassprop1}
\end{split}
\end{align}
where $p=i\sqrt{N^2+m^2}$.
For $z \rightarrow \infty$
\begin{align}
  G_M& (l,m^2) \sim ({1 \over z})^{N-ip}
\label{slargelprop}
\end{align}

As in the case for the graviton, we define,
\be
  G_H (l,\Delta) = ({1 \ov z})^{\Delta} F(\Delta,-N+{1 \ov 2}+\Delta;-2N+1+2\Delta;{1 \ov z})
\label{smassprop}
\ee
We note that,
\be
  G_H (l,\Delta) \sim e^{-\Delta l}
\label{sscaling}
\ee
for non-problematic $\Delta$ and that 
\be
 G_H (l,\Delta) \propto G_M (l,\Delta(\Delta-2N))
\ee
for $\Delta > N$.

As in the graviton case, $G_H (l,\Delta)$ is singular when $\Delta=N-n$ for positive integer $n$.
Following the exact same steps taken in appendix \ref{ap:singprop} we see that,
\begin{align}
\begin{split}
 G_H (l,N-ip) = &{K_{s,n} \ov p+in} G_H (l,N+n) + H_0 (l,N-n) + H_1 (l,N-n) \\
                &+ \OO((p+in)^2)
\end{split}
\end{align}
where $H_0$ and $H_1$ at large $l$ behave as,
\begin{align}
 H_0 (l,N-n) &\sim e^{-(N-n)l} \\
 H_1 (l,N-n) &\sim le^{-(N-n)l}
\end{align}

\subsection{Analytic Continuation}

The sum (\ref{Shateq}) may be expressed as,
\begin{align}
\begin{split}
 \hat{G} (X,X', \mu)= \int_{C_{s1}} \frac{dp}{2\pi i} &{\Ga(-ip-N+1)\Ga(ip+N) \over (-1)^{-ip-N}} \\
                                    & \times G^s_p (X,X') W_{(p)} (\mu) 
\end{split}
\end{align}
where the contour $C_{s1}$ is defined to be one that comes down from $i \infty$ on the left side of the imaginary
axis of the complex $p$ plane, and pivots around $p=iN$ to go back to
$i \infty$ by the right side of the imaginary axis.

Plugging in (\ref{Gpfinal}) in to this equation we obtain,
\begin{align}
\begin{split}
 \hat{G} (X,X', \mu)= \int_{C_{s1}} \frac{dp}{4 \pi p} &
  {\Ga(-ip-N+1)\Ga(ip+N) \over (-1)^{-ip-N}} \\
  & \times ( e^{ip \delta X} + \RR(p) e^{-ip \bar{X}}) W_{(p)} (\mu) \\
  + A e^{N \bar{X}}+ B e^{N \bar{X}} &\bar{X} + C e^{N \bar{X}} {\partial \ov \partial p} W_{p}(\mu) |_{p=iN}
\end{split}
\end{align}
where the additional term comes from the double pole arising from the
additional term in $G_{iN}$ given in (\ref{GiN}).
Note that $W_{iN}(\mu)$ is constant.

We focus our attention on the integral of latter term(+the residual terms),
where the first term just gives the scalar propagator in flat space.
By essentially the same arguments given in the spin 2 case,
the contour of integration for the latter terms can be safely deformed to
the contour $C_{s}$, which we define to run along the real axis of the $p$ plane, with a `jump'
just under $p=iN$. We get,
\begin{align}
\begin{split}
 \hat{G}^{\bar{X}} (X,X', \mu)= \int_{C_{s}} \frac{dp}{4 \pi p} &
  {\Ga(-ip-N+1)\Ga(ip+N) \over (-1)^{-ip-N}} \\
  & \times \RR(p) e^{-ip \bar{X}} W_{(p)} (\mu) \\
  + A e^{N \bar{X}} + B e^{N \bar{X}} &\bar{X} + C e^{N \bar{X}} {\partial \ov \partial p} W_{p}(\mu) |_{p=iN}
\end{split}
\end{align}

After the analytic continuation,
\begin{align}
  X = T + i \frac{\pi}{2},~~~ \mu = il
\end{align}
we finally obtain,
\begin{align}
\begin{split}
 G^{\bar{T}} (T,T', l) = C_{s0} \int_{C_{s}} dp &
 \Ga(ip+N)\Ga(-ip+N) \RR e^{-(N+ip) \bar{T}} Y_{(p)} (il)  \\
 + A' + B' &\bar{T} + C' {\partial \ov \partial p} Y_{p}(il) |_{p=iN}
\label{SGreenfinalp}
\end{split}
\end{align}
where we have conveniently defined,
\be
Y_{(p)} (il) \equiv F(N+ip,N-ip;N+{1\ov2};1-z)|_{z=\cosh^2 {l \over 2 }}
\ee
and we have gotten rid of the hat on the propagator by multiplying $e^{-N \bar{T}}$.

\subsection{A Gauge Argument}

Let's examine the terms,
\be
 A' + B' \bar{T} + C' {\partial \ov \partial p} Y_{p}(il) |_{p=iN}
\ee
of (\ref{SGreenfinalp}).

The first two terms,
$A'+B'T_1 + B'T_2$ vanish when we take derivatives with respect to
both points showing up in the two point function.
In other words, these terms are pure gauge.
Getting rid of this term we can write,
\begin{align}
\begin{split}
 G^{\bar{T}} (T,T', l) = C_{s0} \int_{C_s} dp &
 \Ga(ip+N)\Ga(-ip+N) \RR e^{-(N+ip) \bar{T}} Y_{(p)} (il) \\
 &+ K_{sN} {\p \ov \p p} Y_{(p)} (il)|_{p=iN}
\label{SGreenfinal}
\end{split}
\end{align}

Note that
\be
 I_{sN} \equiv K_{sN} {\p \ov \p p} Y_{(p)} (il)|_{p=iN} \sim l
\ee
for large $l$.

\subsection{The Large $l$ Limit} \label{ss:massless}

Due to the identity between hypergeometric functions,
\be
 Y_p (il) =  { \Ga(N+ {1 \over 2}) \Ga(-2ip)  \over \Ga(N-ip) \Ga({1 \over 2}-ip)} G_H (l,N+ip)
               + { \Ga(N+ {1 \over 2}) \Ga(2ip)  \over \Ga(N+ip) \Ga({1 \over 2}+ip)} G_H (l,N-ip)
\ee
hence the first term in (\ref{SGreenfinal}) can be written as,
\begin{align}
\begin{split}
 \hat{G}^{\bar{T}} = C_{s0} \int_{C_s} dp \RR e^{-(N+ip)\bar{T}}
 [&{\Ga(-ip) \Ga(ip+N) \over 2^{-2ip-1/2}} G_H (l,N+ip) \\
  +&{\Ga(ip) \Ga(-ip+N) \over 2^{2ip-1/2}} G_H (l,N-ip)]
\end{split}
\label{Supdown}
\end{align}

Define the contour $C_{s-}$ to be the contour coming from $-i \infty$ on the left side of the imaginary
axis, pivoting just under $p=iN$ and going back down to  $-i \infty$ on the right side of the imaginary axis.
Define the contour $C_{s+}$ to be the contour coming from $i \infty$ on the left side of the imaginary
axis, pivoting around $p=iN$ and going back up to  $i \infty$ on the right side of the imaginary axis.
Then we may deform the contour of integration for each term to be,
\begin{align}
\begin{split}
 G^{\bar{T}} = 
  &C_{s0} \int_{C_{s-}} dp \RR e^{(-N-ip)\bar{T}}
   {\Ga(-ip) \Ga(ip+N) \over 2^{-2ip-1/2}} G_H (l,N+ip) \\
  +&C_{s0} \int_{C_{s+}} dp \RR e^{(-N-ip)\bar{T}}
   {\Ga(ip) \Ga(-ip+N) \over 2^{-2ip-1/2}} G_H (l,N-ip) \\
   \equiv &I_{s-} + I_{s+}
\end{split}
\label{updown}
\end{align}
 
The poles of the integrand of $I_{s+}$ are given as the following.
\begin{enumerate}
    \item $p=in$ for integers $n$.
    \item $p=-i(N+n)$ for non-negative integer $n$.
    \item The poles of $\sR$.
\end{enumerate}
The only feature we should pay attention to is that
$p=iN$ is a double pole for even dimensions.
All other poles that contriubte are all simple poles.

The poles of the integrand of $I_{s-}$ are given as the following.
\begin{enumerate}
    \item $p=in$ for integers $n$.
    \item $p=i(N+n)$ for non-negative integer $n$.
    \item The poles of $\sR$.
\end{enumerate}
The poles that contriubte will in general all be simple poles.

We can finally write out,
\begin{align}
\begin{split}
 I_{s-}+I_{s+} + I_{sN}
 = &\sum_{n=1}^{\infty} A_{sn} e^{(-N+n)\bar{T}} G_H (l, N+n) \\
 + &\sum_{n= -\infty}^{0} B_{sn} e^{(-N+n)\bar{T}} G_H (l, N-n) \\
 + &\sum_{ia_n :~\text{all poles of }\RR } C_{sn} e^{(-N+a_n)\bar{T}} G^{ij}_{H~i'j'} (l, N-a_n) \\
 + &\delta_{N,[N]} (D_{sN} \bar{T} G_H (l,2N) + E_{sN} {\p \ov \p \Delta} G_H (l, \Delta) |_{\Delta=2N}) \\
 + &K_{sN} {\p \ov \p p} Y_{(p)} (il)|_{p=iN}
\label{sfin}
\end{split}
\end{align}

Note that at large $l$,
\be
 {\p \ov \p \Delta} G_H (l, \Delta) |_{\Delta=2N} \sim l e^{-2Nl}
\ee

\subsection{The Accidental Double Pole} \label{ss:accdp}

One thing we must mention about the
expression (\ref{sfin}) for the scalar propagator is that
the double pole $p=iN$ that arises is
purely a coincidence coming from our assumption that 
the scalar is massless on both sides of the bubble wall.
There is no reason that this should be the case in general
for minimally coupled scalars.

One minimally coupled scalar we know that exists in our
model is the scalar field $\phi$, namely the modulus field.
In the case of this field,
it is certainly natural to assume a mass at least in
the false vaccum. This would modify (\ref{Seq}) so that
$U(X) \rightarrow U(X) + m^2 a(X)^2 \Theta(X-X_0)$.
$G_p(X,X')$ used in the sum (\ref{Seq}) would have to be modified.
If we assume the scalar to be massless in the true vacuum,
it would still be of the form (\ref{Gpfinal})
but the reflection coefficient, $\RR (p)$ would be modified.
In fact, as pointed out in \cite{Freivogel:2006xu},
this shifts the pole at $p=iN$ to $p=i(N-\epsilon)$
where $\epsilon >0$.

Hence in general, the expression (\ref{sfin}) would be
modified to
\begin{align}
\begin{split}
 G^{\bar{T}}
 = &\sum_{n=1}^{\infty} A_{sn} e^{(-N+n)\bar{T}} G_H (l, N+n) \\
 + &\sum_{n= -\infty}^{0} B_{sn} e^{(-N+n)\bar{T}} G_H (l, N-n) \\
 + &\sum_{ia'_n :~\text{all poles of }\RR' } C_{sn} e^{(-N+a'_n)\bar{T}} G^{ij}_{H~i'j'} (l, N-a'_n) \\
\label{sfin1}
\end{split}
\end{align}

Note that for the graviton case, nothing of this sort happens;
the graviton is massless on both sides of the wall.
The reflection coefficient $\RR(p)$ is given exactly by
(\ref{rcoeff}) rendering the pole at $p=iN$
to be at least doubly degenerate.
Unlike for the case of the scalar that provides the tunneling,
the logarithmic piece seems to be a crucial element
of the graviton propagator.

\section{Speculation and Outlook} \label{s:sum}

\subsection{Holographic Correspondence}

For the moment, let's be optimistic and assume that
an $AdS/CFT$ like correspondence exists for a
bulk theory in the flat time-like region of the $D$
dimensional CDL background and the $S^{D-2}$ boundary
at spacelike infinity.
In this section, we will try to make some suggestions
of what such a theory would look like.

For the sake of simplicity of argument,
let's assume the scalar mass is zero on both sides
of the wall.
This is because we don't want to introduce
a mass scale other than the size of the wall,
which comes from the geometry of the background.

FSSY suggested in \cite{Freivogel:2006xu} that
in the 4D case the field theory in
the time-like flat bulk corresponds to a Liouville theory
on the $S^2$ boundary.
In the process they have identified the time coordinate with
the Liouville field of the boundary($L=2T$).
In that sense, we can view time
being emergent from a Liouville field.

We can certainly see something similar in general dimensions.
By writing out the two point functions as we have,
(more precisely, by arranging the terms according to the scaling behavior
with respect to $e^{\bar{T}}$,)
we see that the two point functions(both for the spin 2 and 0 case)
can be basically written as a sum of three kinds of terms,
\begin{align}
 &e^{-(N+n)T_1} e^{-(N+n)T_2} G_H (l,N+n) \quad n\text{ : non-negative integers} \\
 &e^{-(N-a_n)T_1} e^{-(N-a_n)T_2} G_H (l,N-a_n) \quad  ia_n\text{ : poles of }\sR \\
 &e^{-2N\bar{T}} e^{(N+n)T_1} e^{(N+n)T_2} G_H (l,N+n) \quad  n\text{ : non-negative integers}
\end{align}
where $G_H (l,\Delta)$ is a dimension $\Delta$ propagator
with a given spin on $H^{D-1}$.
(There are terms that certainly don't fit in to this framework, 
and we will discuss them later.)
If we assume the existence of a holographic duality of
a field theory in this background,
it is very tempting to view the time $T$
as a dilatonic field on $\Sigma$
by writing the propagator out this way.

Indeed, if we take a slice of our space,
$(T(x),x)$ where $x=(\vec{x},z)$
are the Poincare coordinates on $H^{D-1}$,
the propagator restricted to this slice can be written as a sum of
\begin{align}
 &e^{-(N+n)T(x_1)} e^{-(N+n)T(x_2)} G_H (x_1,x_2,N+n) \\
 &e^{-(N-a_n)T(x_1)} e^{-(N-a_n)T(x_2)} G_H (x_1, x_2,N-a_n) \\
 &e^{-2N\bar{T}} e^{(N+n)T(x_1)} e^{(N+n)T(x_2)} G_H (x_1,x_2,N+n)
\end{align}
If we take these to $\Sigma$ by taking $z \rightarrow 0$
and stripping away the $z$ dependence by defining $T(\vec{x}) \equiv \lim_{z \to 0} T(\vec{x},z)$ we get,
\begin{align}
 &{e^{-(N+n)T(\vec{x_1})} e^{-(N+n)T(\vec{x_2})} \ov |\vec{x_1} -\vec{x_2}|^{2(N+n)} } (t^{ij}_{~~i'j'})\\
 &{e^{-(N-a_n)T(\vec{x_1})} e^{-(N-a_n)T(\vec{x_2})} \ov |\vec{x_1} -\vec{x_2}|^{2(N-a_n)} } (t^{ij}_{~~i'j'})\\
 &(e^{-2N\bar{T}}) {e^{(N+n)T(\vec{x_1})} e^{(N+n)T(\vec{x_2})} \ov |\vec{x_1} -\vec{x_2}|^{2(N+n)} } (t^{ij}_{~~i'j'})
\end{align}
where $t^{ij}_{~~i'j'}$ given by (\ref{deften})
is multiplied to each scalar part for the tensor two point function.
By (\ref{tenssurv}), only the components with indices in the
tangential directions survive at the boundary.
$t^{ij}_{~~i'j'}$ actually is
proportional to that given in equation (2.18)
of \cite{Erdmenger:1996yc}.
We can see that the first two terms are of the same form as two point functions
of (quasi-)primary operators of a CFT given in \cite{Erdmenger:1996yc}
in a dilatonic background $2T(\vec{x})$, and the last with $-2T(\vec{x})$ multiplied by
an additional prefactor.

What these two kinds of propagators mean is not clear,
but it is possible that the graviton and scalar
field correspond to a sum of spin 2 and spin 0 operators
living on the boundary with definite scaling dimensions.

One imaginable scenario is that we have 2 CFTs, $CFT_1$ and $CFT_2$
coupled to possibly a gravity theory such that the action is given by,
\be
 \int \LL_1 (\Omega_1 = e^{2T}) + \int \LL_2 (\Omega_2 = e^{-2T})
\ee
Where $\LL_i (\Omega_i)$ denotes the $CFT_i$ lagrangian with local scaling $\Omega_i$.
This is due to the fact that
we have two distinguishable contributions to our propagator:
the waves going toward the boundary wall
and the waves coming from the boundary wall.
If our bulk field corresponds to an operator sum,
\begin{equation}
 \phi \rightarrow \OO \equiv \sum_{\Delta_1} \OO_1(\Delta_1) + \sum_{\Delta_2} e^{-2NT}\OO_2(\Delta_2)
\label{conjsum}
\end{equation}
with $\OO_1$ being primary operators in $CFT_1$ and
$\OO_2$ being primary operators in $CFT_2$,
the two point function of $\OO$, with fixed $T(\vec{x})$
would indeed look like something we have.\footnote{An alternative
interpretation is offered in \cite{Susskind:2007pv}
where it is conjectured that there is only one CFT and
$\OO_1$ and $\OO_2$ are interpreted as renormalization invariant(``proactive")
and renormalization covariant(``reactive") operators.}

A few comments are to be made.
Trying to interpret the two point function this way,
we notice that we have operators that aren't of dimension $N+n$,
namely ones with dimension $N-a_n$ where $a_n$ depends on
the bubble wall position.
(More precisely put, $a_n$ are real poles of the function,
\be
{F(-N+1,N+1;1+a_n;t) \ov F(-N,N;1+a_n;t)}
\ee
for $t={e^{-X_0} \ov 2\cosh X_0}$.)
This means that we have operators with anomalous
dimensions, depending on a tunable parameter of the theory, $X_0$.
If we give a mass to the scalar, the terms showing up in the
scalar propagator would depend on the mass as well.
But the point is that we have a dimensionful parameter
coming from the geometry of the background,
and that the anomolous dimensions of operators are
related to this by an analytic function.

Also, thinking of graviton fields
on the boundary as dimension 0 operators,
we have a natural interpretation for the logarithmic term.
As we can see from the terms showing up in
the expansion for the graviton propagator
written out in section \ref{ss:sumgrav},
it can be written out as a sum of propagators
that are well behaved at the boundary,
plus a logarithmic(dimension zero) piece.
We've seen in section \ref{ss:log} that
this piece has a fluctuation the size of
the background curvature.
This suggests that the boundary theory
should have geometric fluctuations,
which indeed is coherent with the conjecture
that $T$ is emergent from a dilatonic field
on the boundary theory.
Actually, to stretch our conjecture a bit more,
it is possible that $CFT_1$ mentioned above contains
gravity where the fluctuation of $T$ corresponds to the dilaton.
All such speculation is coherent with the two point function
we have obtained, but much more evidence would be needed to
back up this proposal.

We also note that the coefficients showing up for the three
kinds of propagators in the propagator sum depend on the
reflection coefficient, and in the thin wall limit,
ultimately on the bubble wall position.
If we assume that indeed our bulk fields correspond to a
sum of operators on the boundary,
then how they are summed to give a corresponding bulk field
is dependent upon the bubble wall position.

Another issue we must address are the irregular correlators
that show up for operators of dimension,
$\Delta = (D-2), (D-3)$. These can be seen in equations,
(\ref{eobstr2}) and (\ref{oobstr2}).
The propagator corresponding to $\Delta=(D-3)$ is easy to think about.
In the even dimensional case, they just are propagators of operators
of dimension $(D-3)$. In the odd dimensional case, the leading order behavior
is of dimension $(D-1)$ ($\sim z^{D-1}$,) which doesn't match its scaling dimension
with respect to $T$. We don't quite understand this piece and will ignore it,
as it disappears faster than it should as $z \rightarrow 0$.
Under this prescription, $N+(N-1)=(D-3)$ is not a special case.
In even dimensions, $(N-1)$ is an integer, so it is natural for an
$N+(N-1)$ dimensional operator to show up in the sum.
In odd dimensions, $(N-1)$ is not an integer, so
an $N+(N-1)$ dimensional operator doesn't show up in the sum.

The interpretation of the dimension $(D-2)$ piece seems to
be trickier. Just as with the $(D-3)$ dimensional
propagators, let's choose to discard the pieces with
leading order behavior $\sim z^D$.
Then, if we try to interpret it as
a stress energy tensor as we have suggested in section \ref{ss:set},
we see that it only exists for $CFT_2$, and in the even dimensional case,
is obstructed by a logarithmic term.
The lograrithmic term causes a problem because it
renders the stress energy tensor to be non-transverse.
How to treat this is not entirely clear at the moment.
This is because we have a dimension
zero operator in $CFT_1$ with the same $e^{\bar{T}}$ power
as the stress energy tensor of $CFT_2$.
It would be comforting if we could just get rid of the logarithmic
term by claiming that it comes from
the dimension zero operator and ignore it,
but at the moment it stands as a term we have to deal with.

Also, the fact that a dimension $(D-2)$ operator doesn't show up for
$CFT_1$ is interesting. We have conjectured that gravity would live in
$CFT_1$, so it might be that only $CFT_1$ respects the full diffeomorphism
invariance rendering $T^{ij}_1 =0$, and $CFT_2$ only responds to dilatonic
fluctuations.

There is a different interpretation of this from the framework
of \cite{Susskind:2007pv}. In this case, there is only one stress energy
tensor for the theory in the first place.
The existence of a non-zero stress energy tensor will
be an indication that the Liouville field has decoupled from the
rest of the theory at some fixed point.

As we have already mentioned, the boundary
theory has a tunable parameter: the bubble wall position.
We have seen that the bubble wall position
determines the correspondence between fields
and operator sums.
It also determines the dimensions of operators
that come from the pole of the reflection
coefficient. We have seen in section \ref{ss:oe}
that this is conspicuous in odd dimensions,
as the reflection coefficient has an infinite number
of poles in this case.
Tuning the bubble wall position also seems to trigger some kind of
phase transition in odd dimensions,
as nothing of the sort happens in even dimensions.

This may be attributed to the fact that
for a CDL instanton solution, only the bubble size ${1 \ov \cosh X_0}$
is specified \cite{Coleman:1980aw}.
That is, if $X_0 = a (>0)$ is a good instanton solution,
so is $X_0 = -a$.
The only difference between the two solutions is that the
former has a smaller portion of $dS$ in it.
If we assume some kind of duality between the field theories
with the two instanton solutions as their backgrounds,
$X_0 = 0$ would be a fixed point of the theory.
Why this stands out only in odd dimensions
is not clear at the moment.

\subsection{Outlook}

Although the graviton propagator written out in
section \ref{ss:sumgrav}
and the scalar propagator written out in section 
\ref{ss:accdp} doesn't provide any conclusive evidence
of a holographic duality of two theories we can
expect to fathom, assuming the latter certainly
gives rise to many exciting possibilities.

If indeed such a correspondence were established,
we will be able to gain a route to access
a very novel kind of field theory;
that is, one on Euclidean space with
two CFTs (one possibly containing gravity)
coupled in a rather peculiar way.
This theory would have a tunable parameter, and might
have a phase transition in odd dimensions.

\section*{Acknowledgements}

Thanks to Hong Liu for introducing me to this subject and
providing direction and support throughout the writing of this paper.
I also thank Leonard Susskind and Yasuhiro Sekino for
providing valuable insight on this subject and
discussions on various aspects of this calculation,
Ben Freivogel for being kind enough to share his time
for discussions during his visit to MIT, and John McGreevy for
many useful discussions and a wonderful string theory course
that was very helpful to this work.
This work was supported in part by funds provided by the U.S. Department of Energy
(D.O.E.) under cooperative research agreement DE-FG0205ER41360,
and partly by the Korea Foundation for Advanced Studies.

\appendix

\section{The asymptotic behavior of $\sR$ for odd dimensions} \label{ap:poles}

In order to examine the the poles of $\sR$ in the limit $k \rightarrow - i\infty$
it is convenient to consider the asymptotic behavior of $\sin \pi x F(-N,N,1+x,t)$
in the limit $x \rightarrow - \infty$ where we have cancelled
all the poles of the hypergeometric function by the multiplication of the sine function.
This is because we are interested in the imaginary poles of $F(-N+1,N+1;1-ik;t)/F(-N,N;1-ik;t)$
for $k \rightarrow - i\infty$ and we know that the denominator gets rid of the poles,
$ik=\text{integer}$ coming from the numerator, and hence our interest lie
in the zeros of $\sin \pi x F(-N,N,1+x,t)$.

We use the relations,
\begin{align}
  F(a,b;c;z) &= (1-z)^{c-a-b} F(c-a,c-b;c;z) \\
  \Ga(z) \Ga(1-z) &=  \pi \csc \pi z
\end{align}
and
\begin{align}
\begin{split}
  F(&a,b;c;z) = \\
  &{\Ga(c) \Ga(c-a-b) \ov \Ga(c-a) \Ga(c-b) } F(a,b;a+b-c+1;1-z) \\
  &+ (1-z)^{c-a-b} {\Ga(c) \Ga(a+b-c) \ov \Ga(a) \Ga(b)} F(c-a,c-b;c-a-b+1;1-z)
\end{split}
\end{align}
to obtain
\begin{align}
\begin{split}
 & \sin \pi x F(-N,N;1+x;t) = \\
 & {\Ga(- \! x \! - \! N) \Ga(- \! x \! + \! N) \ov \Ga(-x)^2 } [ ({t \ov 1 \! - \! t})^{-x} ({1 \! - \! t \ov -x})
  N  \sin \pi N F(1 \! + \! N \! , 1 \! - \! N \!  ; 1-x;t) \\
 &+ \sin \pi x F(-N,N;-x;1-t)   ]
\end{split}
\end{align}

First note that for $x \rightarrow - \infty$
\begin{align}
  \Ga (-x + a )\Ga (-x - a ) / \Ga(-x)^2 \approx 1
\end{align}
for any fixed real number $a$. Also in this limit,
\begin{align}
  F(a,b;-x;z) = 1 + \OO ( {1 \ov |x|} )
\end{align}
so up to leading order in $1/|x|$ we get,
\begin{align}
  \sin \pi x F(-N,N;1+x;t) \approx
  ({t \ov 1-t})^{-x} ({1-t \ov -x}) N \sin \pi N + \sin \pi x
\end{align}
In the case $t/(1-t) \leq 1$ we see that the first terms in this equation
vanishes in the desired limit. For $t/(1-t) > 1$, the second term becomes irrelevant.

Hence we can write the asymptotic behavior for our function in the limit $x \rightarrow -\infty$
as the following.
\begin{align}
  \sin \pi x F(-N,N;1+x;t) \approx
\begin{cases}
  [(1-t)N \sin \pi N] {(t/(1-t))^{-x} \ov -x}  \quad  t > 1/2 \\
  \sin \pi x  \quad  t \leq 1/2
\end{cases}
\end{align}

Note that we expect an infinite number of real zeros in $x$ of
$F(-N,N;1+x;t)$ for $t \leq 1/2$ where for $t > 1/2$ the number of
real zeros becomes finite.

Now note that since $\sR$ is analytic for general $t$, for a given
neighborhood of such $t$, the number of poles should be the same.
Since $\sR$ has an infinite number of imaginary poles as
$k \rightarrow - i\infty$ for $t \leq 1/2$,
we know that the number of poles of $\sR$ in the lower half plane of $k$
should be infinite for a given neighborhood around $t=1/2$.
But we now also know from the asymptotic behavior of $\sin \pi x F(-N,N,1+x,t)$
that $\sR$ has a finite number of poles on the lower imaginary axis.
Hence $\sR$ has an infinite number of poles that aren't imaginary in
the lower half plane for $1/2<t<1/2+\epsilon$ for some $\epsilon>0$.

One might question the validity of this argument by
questioning the statement that $\sR(ix)$ has an infinite number of
real poles at $t=1/2$.
Since $t = 1/2$ is a marginal value,
one might feel that the argument based on the $x \rightarrow -\infty$
behavior of the function might not hold up.
That is, it is possible that as
$t \rightarrow 1/2-$, $x_M>0$ for which at $x<-x_M$
we may safely approximate $\sin \pi x F(-N,N;1+x;t) \approx \sin \pi x$
might tend to infinity which would render the previous argument invalid.

Fortunately, we can explicitly prove that $\sR(ix)$ has an
infinite number of real poles for $t=1/2$,
which goes like the following.
Let's deal with $\sR$ directly for simplicity.

We write,
\be
 \sR(ix) =  {N(1-t)F(-N+1,N+1;1+x;1/2) \ov (x-N)F(-N,N;1+x;1/2)}
\ee
For sake of convenience, we will prove the equivalent statement that,
\be
 f(x) \equiv  {N \ov 4}{F(-N+1,N+1;1+x;1/2) \ov F(-N,N;1+x;1/2)}
\ee
has an infinite number of real poles.

Using the equalities,
\begin{align}
\begin{split}
 F(-N+1,N+1;1+x;1/2)=&{x+N \ov N}F(-N,N+1;1+x;1/2) \\
                    -&{x-N \ov N}F(-N+1,N;1+x;1/2) \\[10pt]
 F(-N,N;1+x;1/2)=&{1 \ov 2}F(-N,N+1;1+x;1/2) \\
                +&{1 \ov 2}F(-N+1,N;1+x;1/2)
\end{split}
\end{align}
and
\be
 F(a,1-a;1+x;1/2)=2^{-x}\pi^{1/2} {\Ga(1+x) \ov \Ga({1 \ov 2}a+{1 \ov 2} x +{1 \ov 2}) \Ga(-{1 \ov 2}a+{1 \ov 2} x +1) }
\ee
we get,
\be
 f(x) \equiv  { {1 \ov \Ga(a+1/2) \Ga(a+N)} - {1 \ov \Ga(a) \Ga(a+N+1/2)} \ov {1 \ov \Ga(a+1/2) \Ga(a+N+1)} + {1 \ov \Ga(a+1) \Ga(a+N+1/2)}}
\ee
where $a= {1 \ov 2} (x-N) $.
Note that written in this way, both the numerator and denominator are analytic functions with no poles in the $x$ plane.
In order find the poles of $f(x)$, all we have to do is find the zeros of the denominator that aren't cancelled by a
zero of the numerator.

Define the function,
\be
 g(a) \equiv  {\Ga(a) \ov \Ga(a+1/2)}
\ee
Then the zeros of the numerator come from the equation,
\be
 g(a) =  g(a+N)
\ee
and the zeros of the denominator come from,
\be
 g(a+1/2) =  -g(a+N+1/2)
\ee

\begin{figure}[!ht]
\leavevmode
\begin{center}
\epsfysize=6cm
\epsfbox{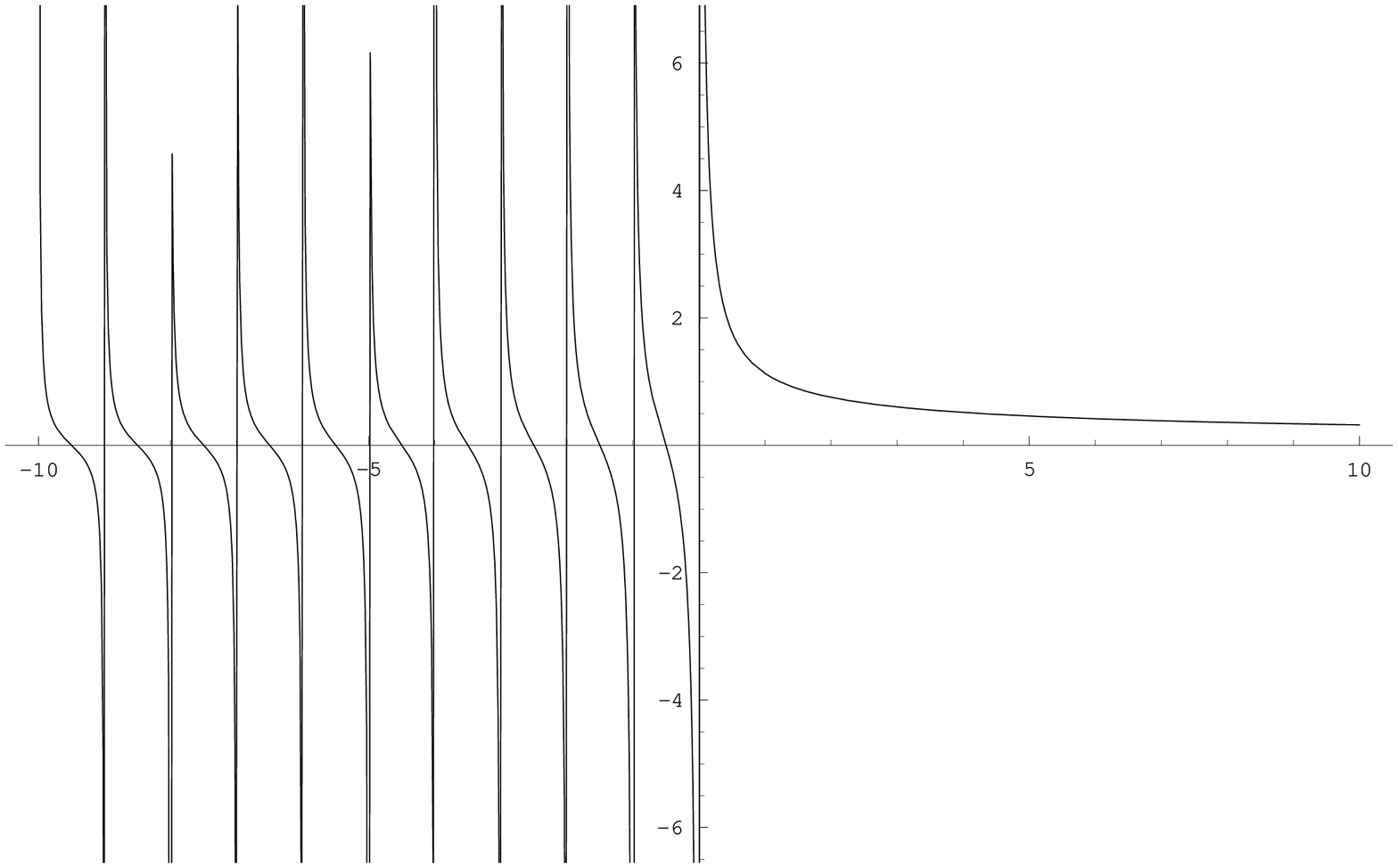}
\end{center}
\caption{\small The plot $a$ vs. $g(a)$.
}
\label{g(a)}
\end{figure}

From the analytic property of $\Ga(a)$, we can infer that of $g(a)$.
To sum up, $g(a)$ has the following properties.
\begin{enumerate}
    \item For $a>0$, $g(a)$ monotonically decreases from $+\infty$ at $a=0+$ to $0$ as $a \rightarrow \infty$.
    \item For negative integer $n$, $g(a)$ monotonically decreases in the interval $(n,n+1)$ from
          $g(n+0) \rightarrow \infty$ to $g(n+1-0) \rightarrow -\infty$.
    \item For negative integer $n$, $g(n+1/2) =0$.
\end{enumerate}
These facts are evident in figure \ref{g(a)}.

Hence for half integer $N=M+1/2$, there are $M$ roots to $g(a) =  g(a+N)$ each in the interval,
$(n,n+1/2)$ for $n=-1, \cdots, -M$.

Let's get to $g(b) =  -g(b+N)$.
First of all, there is one root in each interval $(n+1/2,n+1)$ for negative integer $n$.
Also, there is one root in each interval $(n,n+1/2)$ for integer $n<-M$.

Translating this for $a=b-1/2$, the roots are given as the following.
\begin{enumerate}
    \item There is one root in each interval $(n,n+1/2)$ for negative integer $n$.
    \item There is one root in each interval $(n-1/2,n)$ for integer $n<-M$.
\end{enumerate}

Hence we see that there are an infinite number of
negative zeros appearing in every $1/2$ length interval in the denominator that aren't
cancelled by zeros of the numerator.
This completes the proof.

\section{The Explicit Expression for $w^I(\alpha_p)$ and $Q_p$} \label{ap:Wgrav}

Defining $\alpha_p$ as,
\begin{align}
	& \alpha_p (z) = F (\frac{D+2}{2} +ip, \frac{D+2}{2} -ip ; \frac{D+3}{2} ; 1-z )
\label{defalph}
\end{align}
we get,
\begin{align}
\begin{split}
  w^1(\alpha_p) =& \frac{4(D-2)}{D(D-3)}\big[\big(p^2 - \frac{3}{4} D^2+2D +1 \big)z(z-1)- {D(D-3) \ov 4}\big] \alpha_p(z) \\
                      & + \frac{8(D-2)}{D(D+3)} \big(p^2 +  (\frac{D+2}{2})^2 \big)z(z- \frac{1}{2})(z-1) \beta_p(z) \\[10pt]
  w^2(\alpha_p) =& (1-z) \big[ \frac{2(D-2)^2}{D(D-3)} \big(p^2 - \frac{3}{4} D^2+2D +1 \big)z+ (D-1)(D-2) \big] \alpha_p(z) \\
                      & - \frac{4(D-2)^2}{D(D+3)} \big(p^2 +  (\frac{D+2}{2})^2 \big)z(z-1)(z- \frac{D-1}{D-2}) \beta_p(z)\\[10pt]
  w^3(\alpha_p) =& \big[-\frac{2(D-2)^2}{D(D-3)} \big(p^2 - \frac{3}{4} D^2+2D +1 \big)z(z-1)+ \frac{(D-1)(D-2)}{2} \big]
                                                 \alpha_p(z) \\
                      &  - \frac{4(D-2)^2}{D(D+3)} \big(p^2 +  (\frac{D+2}{2})^2 \big) z(z- \frac{1}{2})(z-1) \beta_p(z) \\
\end{split}
\label{wIp}
\end{align}
where $\beta_p$ is defined as,
\be
 \beta_p (z) = -{ (D+3)/2 \over (N+2)^2 + p^2} {d \alpha_p (z) \over dz}
\label{defbet}
\ee

Also,
\begin{align}
\label{Q_p}
  Q_p =&  [{i\Ga (D/2) D(D-3) \over 4 \pi^{D/2} (D-2)! (D-2)^2 (D^2-1) } ]
	         { p(p^2+(N+1)^2)\Ga(-ip+N-1) \over \Ga(-ip-N+2)} 
\end{align}

\section{Transverse Traceless \\ Tensor Propagators in $H^{(D-1)}$} \label{ap:tpinH}

We wish to examine the traceless spin 2 particle
propagator on $H^{D-1}$ with mass $m$.
Note that we know from $AdS/CFT$ that this taken to the boundary
corresponds to the propagator of symmetric traceless tensor operators
with dimension $\Delta=N+\sqrt{N^2+m^2}$ (see, for example, \cite{Polishchuk:1999nh}.)

The equation for the propagator for a massive traceless spin 2 particle in
$H^{(D-1)}$ can be derived from the action,
\be
 \int d^{D-1} \sqrt{g} ({\bf\rm R} - 2\Lambda + {1 \ov 2} m^2 h^{i j} h_{i j})
\ee
where $g_{i j} = \gamma_{i j} + h_{i j}$ with the $H^{D-1}$
metric $\gamma_{i j}$.
${\bf\rm R}$ is the Ricci scalar for the metric $g_{i j}$
We only focus on the traceless part of the spin 2 tensor for now,
for reasons that will soon be clear.
All indices are raised and lowered by the background metric.

We work with the background curvature radius, $R^2 = -1$.
Then the Ricci scalar of the background is given to be $-(D-1)(D-2)$,
and the cosmological constant would be $-{1 \ov 2}(D-2)(D-3)$.

This can be perturbed to give the equations of motion (\cite{Barth:1983hb}),
\begin{align}
\begin{split}
 \Box h_{i j} &+ g_{i j} \nabla^k \nabla^l h_{k l}
 - \nabla_j \nabla_k h_i^k - \nabla_i \nabla_k h_j^k \\
 &+ 2 R^{~k~l}_{i~j}h_{k l} + 2 R_i^k h_{k j} - R h_{i j}
 + 2 \Lambda h_{i j} = m^2 h_{i j}
\end{split}
\end{align}
where the covariant derivatives, the Ricci tensors/scalar and the Riemann tensors
all are given with respect to the background metric.

The l.h.s. of this equation has zero divergence.
This can be seen by explicit calculation,
or from the Bianchi identity.
Hence for massive tensors,
the transverseness of the propagator would not be a gauge condition,
it would be a constraint coming from the equation of motion.

Using the transverseness of $h_{i j}$, the equation of motion reduces to
\be
 (\Box +2-m^2)h_{i j}=0
\ee
The equation for the propagator can be obtained to be,
\be
 (-\widetilde{\Box}_1-2+m^2) G^{ij}_{M~i'j'} (l(\HH_1, \HH_2),m^2) =
 {1 \over \sqrt{\gamma}}
 (\gamma^{(i}_{(i'} \gamma^{j)}_{j')} - {1 \ov D-1}\gamma^{ij} \gamma_{i'j'} ) \delta (\HH_1, \HH_2)
\ee
with the constraint,
\be
 \nabla_a G^{ij}_{M~i'j'} (l(\HH_1, \HH_2),m^2) =0
\ee
Note that the delta function on the righthand side of the equation for
the propagator is not projected to be transverse, so it is actually
zero for distinct $\HH_1$ and $\HH_2$.

We can solve this by following the steps sketched in \cite{Allen:1985wd}, where first we solve,
\be
 (-\widetilde{\Box}_1-2+m^2) G^{ij}_{M~i'j'} (l(\HH_1, \HH_2),m^2) = 0
\label{bitendef}
\ee
for the maximally symmetric bitensor
but now take the solution most singular at $l=0$ and obtain the multipicative constant by
comparing it to the flat limit.

This can be done via the exact same procedure we obtained $W^{ij}_{(p)i'j'}$,
but we impose different boundary conditions as we are working in a non-compact space.
The solution is,
\be
  G^{ij}_{M~i'j'} (l,m^2) = A(m^2) w^I (a_{i\sqrt{N^2 + m^2}}) t^{ij}_{I~i'j'} |_{z=\cosh^2 {l \over 2 }}
\label{massprop1}
\ee
where we define,
\begin{align}
	a_p (z) = ({1 \over z})^{{(D+2) \over 2} - ip}
	            F (\frac{D+2}{2} -ip, \frac{1}{2} -ip ; 1-2ip ; {1 \over z} )
\end{align}
$A(m^2)$ is some constant and $t^{ij}_{I~i'j'}$ and $w^I (a_p )$ are given by
(\ref{ttensor1}), (\ref{ttensor2}), (\ref{ttensor3}), and (\ref{wIp}).

From the fact that
\begin{align}
	a_p (z) \sim ({1 \over z})^{{(D+2) \over 2} - ip} \quad \text{for }z \rightarrow \infty
\end{align}
we see that for $z \rightarrow \infty$
\begin{align}
  w^I (a_p) t^{ij}_{I~i'j'} \sim ({1 \over z})^{N-ip} t^{ij}_{~~i'j'}
\label{largelprop}
\end{align}


Hence we notice that the scaling dimension of
$w^I (a_{ix}) t^{ij}_{I~i'j'}$ is $\Delta = N+x$.
This can be seen by writing the geodesic length in $H^{D-1}$ in Poincare coordinates.
If we write the metric as,
\be
 ds^2 = {dz^2 + dx_1^2 + \cdots + dx_{D-2}^2 \ov z^2 }
\ee
the length of the geodesic connecting the two points $(z, \vec{x})$ and $(z', \vec{x'})$ is given as,
\be
 \cosh^2 {l \ov 2} = {(z+z')^2 + (x-x')^2 \ov zz'}
\ee
so in the limit, $z, z' \rightarrow 0$,
\be
 w^I (a_{ix}) t^{ij}_{I~i'j'} \sim (\cosh^2 {l \ov 2})^{-N-x} t^{ij}_{~~i'j'}  \sim z^{N+x} z'^{N+x} (x-x')^{-2N-2x} t^{ij}_{~~i'j'}
\ee

We wish to extend the propagator $G^{ij}_{M~i'j'}$ so it could have a general
scaling dimension.
But as can be seen from the expression (\ref{massprop1}),
a massive $H^{D-1}$ propagator with mass $m$ has dimension, $\Delta = N + \sqrt{N^2+m^2}$.
Hence the pieces with dimension $\Delta < N$ can't possibly be written in terms of massive propagators.

Also $A(m^2)$ exhibits singular behavior
(hence forbidding the propagator of having certain scaling dimensions)
if we try to generalize
(\ref{massprop1}) by replacing $i\sqrt{N^2+m^2}$ by $i(\Delta-x)$.
$A(\Delta)$ is evaluated up to a trivial multiplicative factor
explicitly in appendix \ref{ap:coeff}, and we will address relevant issues there.

The important conclusion is that we will define the ``generalized Green function"
\be
  G^{ij}_{H~i'j'} (l,\Delta) = w^I (a_{i(\Delta-N)}) t^{ij}_{I~i'j'} |_{z=\cosh^2 {l \over 2 }}
\label{masspropap}
\ee
that is, as the maximally symmetric bitensor
with definite scaling dimension $\Delta$. We note that,
\begin{align}
\begin{split}
  G^{ij}_{H~i'j'} & (l,\Delta)  \sim C (\!\Delta\!-\!2N\!)(\!\Delta\!-\!2N\!+1) e^{-\Delta l} t^{ij}_{~~i'j'} +
                                   \OO \left( e^{-(\Delta+2) l} \right) \\
  & \sim C (\!\Delta\!-\!2N\!)(\!\Delta\!-\!2N\!+1) {z^{\Delta} z'^{\Delta} \ov |x-x'|^{2\Delta}}  t^{ij}_{~~i'j'}
     + \OO \left( {z^{\Delta+2} z'^{\Delta+2} \ov |x-x'|^{2\Delta+4}} \right) 
\end{split}
\label{gscalingap}
\end{align}
for all non-problematic(we will shortly explain what we mean by `problematic') $\Delta$.
Also,
\be
  G^{ij}_{H~i'j'} (l,\Delta) \propto G^{ij}_{M~i'J'} (l,\Delta(\Delta-2N))
\ee
for $\Delta > N,~\Delta \neq 2N$.

One thing we must note is that $G^{ij}_{H~i'j'} (l,2N)$ is not a
propagator for a spin 2 tensor with $m^2=0$. This is because that the
equation for the transverse traceless massless spin 2 propagator is,
\be
 (-\widetilde{\Box}_1-2) G^{ij}_{M~i'j'} (l(\HH_1, \HH_2),0) =
 {1 \over \sqrt{\gamma}} \delta^{ij}_{~~i'j'} (\HH_1, \HH_2)
\ee
where the delta function on the r.h.s. is a
delta function projected on to transverse-traceless modes,
so it is not zero for distinct $\HH_1, \HH_2$ in general.
This situation arises
because the transverseness of the propagator
doesn't come from the equation of motion and
has to be imposed as a gauge condition.
This propagator is written out in a form compatible with our
formalism in \cite{D'Hoker:1999jc}.

One more thing we have to be concerned about is
that $a_p$(and hence $G^{ij}_{H~i'j'}(l,N-ip)$)
is singular for $1-2ip=-2n+1$ for positive $n$.
We are spared from some worry because in the case, $1-2ip=-2n$
we get,
\be
 a_p (z) = ({1 \over z})^{{(D+1) \over 2} - n}
           F (\frac{D+1}{2} -n, -n ; -2n ; {1 \over z} )
\ee
so the hypergeometric function becomes a polynomial,
stopping short of the divergent piece.
So we just concern ourselves with the case, $p=-in$
for positive integer $n$.

In appendix \ref{ap:singprop} we will show that by expanding
around $p=-in$, we can write,
\begin{align}
\begin{split}
\label{gsing}
 G^{ij}_{H~i'j'}(l,N-ip) &= {1 \ov p+in} K_{-1,n} G^{ij}_{H~i'j'}(l,N+n)  \\
                             &+ H^{ij}_{0~i'j'} (l,N-n) + (p+in) H^{ij}_{1~i'j'} (l,N-n) \\
                             &+ \OO((p+in)^2)
\end{split}
\end{align}
and that for large $l$,	
\begin{align}
 H^{ij}_{0~i'j'} (l,N-n) \sim e^{-(N -n)l} t^{ij}_{~~i'j'}\\
 H^{ij}_{1~i'j'} (l,N-n) \sim le^{-(N -n)l} t^{ij}_{~~i'j'}
\end{align}

\section{The Graviton Propagator in Flat Space} \label{ap:coeff}

The graviton propagator in flat space can be obtained by
\begin{align}
\begin{split}
 \int {d^{D-1}k \ov (2\pi)^{D-1}} {i \ov k^2 + m^2}
 \big( -\delta^{(a}_{(a'} \delta^{b)}_{b')} + {2 \ov D-2} \eta^{ab}\eta_{a'b'}
 - {2(D-3) \ov D-2} {k^a k^b k_{a'}k_{b'} \ov m^4} \\
 + {2 \ov D-2} {k^a k^b \eta_{a'b'} + \eta^{ab} k_{a'} k_{b'} \ov m^2}
 -{ k^{(a} \delta^{b)}_{(a'} k_{b')} \ov m^2} \big)
\end{split}
\end{align}

This can be written in the form (\ref{masspropap}).
For $f(x) \equiv {m^{(D-3)/2} \ov x^{(D-3)/2}} K_{(D-3)/2}(mx)$,
\be
 w^1_{flat} (l) \propto f(l)+{2(D-3) \ov m^4 (D-2)} ({f'(l) \ov l^3} - {f''(l) \ov l^2}) - {4 \ov m^2 (D-2)} {f'(l) \ov l}
\ee
up to a constant independent of mass.
For $l \rightarrow 0$,
\be
 w^1_{flat} (l) \sim  {1 \ov m^4} {1 \ov l^{D+1}}
\ee
up to a constant independent of mass.

Now for $l \rightarrow 0$ since
\be
 a_p (z) \sim {\Ga(1-2ip) \Ga((D-1)/2) \ov \Ga({D+2 \ov 2}-ip) \Ga({1 \ov 2}-ip)} ({1 \ov l})^{(D-1)} 
\ee
we obtain
\be
 w^1_p (l) \sim  {\Ga(1-2ip) \ov \Ga({D+2 \ov 2}-ip) \Ga({1 \ov 2}-ip)} ({1 \ov l})^{(D+1)} 
\ee
up to a constant independent of $p$.

Comparing these two values for a given mass(with $p=i\sqrt{N^2+m^2}$),
we obtain up to a non-singular constant,
\begin{align}
\begin{split}
 A(m^2) &\propto {1 \ov m^4} {\Ga({D+2 \ov 2}+\sqrt{N^2+m^2}) \Ga({1 \ov 2}+\sqrt{N^2+m^2}) \ov \Ga(1+2\sqrt{N^2+m^2})} \\
 &\propto {1 \ov m^4} {\Ga({D+2 \ov 2}+\sqrt{N^2+m^2}) \ov \Ga(1+\sqrt{N^2+m^2})}
\end{split}
\end{align}

Trying to generalize this for a general scaling dimension we get,
\be
 A(\Delta) \propto {1 \ov (\Delta(\Delta-2N))^2} {\Ga(\Delta+2) \ov \Ga(\Delta-N+1)}
\ee
This is singular for $\Delta=0,~2N$ and $-n-1$ for positive integer $n$.
Also note that this is zero for $\Delta=N-n$ for positive integer $n$.
This leads to the interesting fact that due to (\ref{gsing}),
\be
 \lim_{\Delta \to N-n} A(\Delta) G^{ab}_{H~a'b'} (l,\Delta) \propto G^{ab}_{H~a'b'} (l,N+n) 
\ee
for positive integer $n$.

\section{Deconstructing Singular \\ Tensor Propagators} \label{ap:singprop}

We deal with the singularity of $G^{ij}_{H~i'j'}(l,\Delta)$ at $\Delta =N-n$
for positive integer $n$ by writing,
\begin{align}
\begin{split}
 a_p (z) = &({1 \over z})^{{(D+2) \over 2} -ip} f(p,2n-1,z) \\
           &+ {1 \ov p+in} ({i \ov 2}) {({(D+2) \ov 2}-ip)_{2n} ({1 \ov 2}-ip)_{2n} \ov 2n! (1-2ip)_{2n-1}} \\
           &\times ({1 \over z})^{{(D+2) \over 2} -ip+2n}
             F (\frac{D+2}{2} -ip+2n, \frac{1}{2}-ip+2n ; 2n+1 ; {1 \over z} )
\end{split}
\end{align}
where we have defined,
\be
 (x)_n \equiv x(x+1) \cdots (x+n-1)
\ee
and $f(p,m,z)$ is the polynomial
\be
 f(p,m,z) \equiv \sum_{n=0}^m {1 \ov n!} {((D+2)/2-ip)_n (1/2-ip)_n \ov (1-2ip)_n} ({1 \ov z})^n
\ee
To put this in a form which is more useful, we expand the latter part of $a_p$
around $p=-in$ for which we get,
\begin{align}
\begin{split}
 a_p (z) &= {1 \ov p+in} K_{-1,n} a_{in} (z) \\
         & + [({1 \over z})^{{(D+2) \over 2} -n} f_{-in}(z) + K_{0,n} a_{in} (z) + L_{0,n} c_{in} (z)] \\
         & + (p+in) [(I_{1,n} \ln z+J_{1,n}) ({1 \ov z})^{{(D+2) \ov 2} -n} f_{-in} (z) + K_{1,n} a_{in} (z) \\
         & \qquad \qquad +  L_{1,n} c_{in} (z) + M_{1,n} d_{in} (z) ] \\
         & + \OO((p+in)^2) \\
         &\equiv {1 \ov p+in} K_{-1,n} a_{in} (z) + h_{0,-in} (z) + (p+in) h_{1,-in} (z) \\
         &+ \OO((p+in)^2)
\end{split}
\end{align}
Where we have conveniently defined,
\begin{align}
 &f_{-in}(z)  \!  \equiv  \! f(-in,2n-1,{1 \ov z}) \\
 &c_{in} (z)  \! \equiv  \! {\p \ov \p p} ({1 \over z})^{{(D \! + \! 2) \over 2} - \! ip \! + \! 2n}
 F (\frac{ \! D \! + \! 2 \! }{2}  \! - \! ip \! + \! 2n, \frac{1}{2}- \! ip \! + \! 2n ; 2n \! + \! 1 ; {1 \over z} )|_{p=-in} \\
 &d_{in} (z)  \! \equiv  \! {\p^2 \ov \p p^2} ({1 \over z})^{{(D \! + \! 2) \over 2} - \! ip \! + \! 2n}
 F (\frac{ \! D \! + \! 2 \! }{2}  \! - \! ip \! + \! 2n, \frac{1}{2}- \! ip \! + \! 2n ; 2n \! + \! 1 ; {1 \over z} )|_{p=-in}
\end{align}

We can finally write,
\begin{align}
\begin{split}
 G^{ij}_{H~i'j'}(l,N-ip) &= {1 \ov p+in} K_{-1,n} G^{ij}_{H~i'j'}(l,N+n)  \\
                         &+ H^{ij}_{0~i'j'} (l,N-n) + (p+in) H^{ij}_{1~i'j'} (l,N-n) \\
                         &+ \OO((p+in)^2)
\end{split}
\end{align}
where
\begin{align}
 H^{ij}_{0~i'j'} (l,N-n) &\equiv w^I (h_{0,-in} (z)) t^{ij}_{I~i'j'} \label{H0} \\
 H^{ij}_{1~i'j'} (l,N-n) &\equiv w^I (h_{1,-in} (z)) t^{ij}_{I~i'j'} \label{H1}
\end{align}
Note that for $l \rightarrow \infty$ (since $n>0$),
\begin{align}
 h_{0,-in} (z) &\sim e^{-({(D+2) \over 2} -n)l} \\
 h_{1,-in} (z) &\sim l e^{-({(D+2) \over 2} -n)l}
\end{align}
and hence,
\begin{align}
 H^{ij}_{0~i'j'} (l,N-n) &\sim e^{-(N -n)l} t^{ij}_{~~i'j'}\\
 H^{ij}_{1~i'j'} (l,N-n) &\sim le^{-(N -n)l} t^{ij}_{~~i'j'}
\end{align}

\section{Degenerate Modes of the Graviton} \label{ap:degmodes}

In this section, we will identify the degenerate modes of the transverse-traceless
graviton propagator in $H^{D-1}$.

We start with the scalar mode, $E^{(p)}$ such that,
\be
 \widetilde{\Box} E^{(pv)} = -(N^2 +p^2)E^{(pv)}
\ee
In $H^{D-1}$ we find that,
\begin{multline}
 \widetilde{\Box} (\tilde{\nabla}_i \tilde{\nabla}_j- {\tilde{\gamma}_{ij} \ov D-1} \widetilde{\Box})E^{(pv)} = \\
 (-N^2-p^2 -2(2N+1)) (\tilde{\nabla}_i \tilde{\nabla}_j- {\tilde{\gamma}_{ij} \ov D-1} \widetilde{\Box}) E^{(pv)}
\end{multline}

Also,
\begin{align}
 (\tilde{\nabla}^i \tilde{\nabla}_i- {\delta^i_i \ov D-1} \widetilde{\Box}) E^{(pv)} &= 0\\
 \tilde{\nabla}^i (\tilde{\nabla}_i \tilde{\nabla}_j- {\tilde{\gamma}_{ij} \ov D-1} \widetilde{\Box}) E^{(pv)} &=
 {D-2 \ov D-1} (-N^2-p^2 -(2N+1)) \tilde{\nabla}_j E^{(p)}
\end{align}

Hence $(\tilde{\nabla}_i \tilde{\nabla}_j- {\tilde{\gamma}_{ij} \ov D-1} \widetilde{\Box}) E^{(pv)}$ is
a symmetric transverse traceless spin 2 mode for $p=i(N+1)$, and hence its eigenvalue with respect to
$\widetilde{\Box}$ would be $-2N-1$.
Therefore this is degenerate with the spin 2 modes $r^{(pu)}_{ij}$ (whose eigenvalues are
given by $-(N^2+2+p^2)$) with $p=i(N-1)$.

For the vector mode, $F^{(pw)}_i$ such that,
\be
 \widetilde{\Box} F^{(pw)}_i = -(N^2 +p^2+1)F^{(pw)}_i
\ee
we find in $H^{D-1}$,
\be
 \widetilde{\Box} F^{(pw)}_{(i|j)} = (-N^2 -p^2-1-(2N+2))F^{(pw)}_{(i|j)}
\ee

Also since $F^{(pv)}_i$ are transverse,
\begin{align}
 F^{(pv)}_{(i|i)} &= 0\\
 \tilde{\nabla}^i F^{(pw)}_{(i|j)} &= (-N^2 -p^2-1-2N) F^{(pw)}_{j}
\end{align}

So $F^{(pw)}_{(i|j)}$ is
a symmetric transverse traceless spin 2 mode for $p=i(N+1)$, and hence its eigenvalue with respect to
$\widetilde{\Box}$ would be $-2$.
Therefore this is degenerate with the spin 2 modes $r^{(pu)}_{ij}$ (whose eigenvalues are
given by $-(N^2+2+p^2)$) with $p=iN$.

Now let's show that all $r'^{(i(N-1)u)}_{ij}$ come from
$E^{(i(N+1)v)}$
and that all $r'^{((iN)u)}_{ij}$ come from
$F^{(i(N+1)w)}_{i}$ where $r'^{((p)u)}_{ij}$ are defined by (\ref{defZpr}).

Note that by the form of
$(\tilde{\nabla}_i \tilde{\nabla}_j- {\tilde{\gamma}_{ij} \ov D-1} \widetilde{\Box}) E^{(pv)}$,
this has even parity, and hence this certainly cannot saturate $\{ r^{(pu)}_{ij} \}$.
But our objective would be to get rid of the modes contributing to $W^{ij}_{(p)i'j'}$
with $p=iN,i(N-1)$ in our propagator and as will be shown, this can be done.

Define,
\begin{align}
 Z_{(p)} &= \sum_v E^{(pv)\dagger}(\mathcal{H}) E^{(pv)}(\mathcal{H}') \\
 Z^i_{(p)i'} &= \sum_v F^{(pv)i\dagger}(\mathcal{H}) F^{(pv)}_{i'}(\mathcal{H}')
\end{align}
for properly normalized, regular $E^{(pv)}$ and $F^{(pw)}_{i}$.
These are maximally symmetric bitensors as they are invariant under any isometries.
Also, they show regular behavior at $\mathcal{H} = \mathcal{H}'$, i.e. the
coincident point.
A covariant derivative of a maximally symmetric bitensor is also a
maximally symmteric bitensor, hence so are,
\begin{align}
 Z^{ij}_{1(p)i'j'}&=
 (\tilde{\nabla}^i \tilde{\nabla}^j- {\tilde{\gamma}^{ij} \ov D-1} \widetilde{\Box})
 (\tilde{\nabla}_{i'} \tilde{\nabla}_{j'}- {\tilde{\gamma}_{i'j'} \ov D-1} \widetilde{\Box}) Z_{(p)} \\
 Z^{ij}_{2(p)i'j'}&=  Z^{(i|j)}_{(p)(i'|j')}
\end{align}

From the mode sum and by the behavior of the individual modes for $p=i(N+1)$,
$Z^{ij}_{1(i(N+1))i'j'}$ and $Z^{ij}_{2(i(N+1))i'j'}$ are symmetric,
transverse, traceless maximally symmetric bitensors behaving regularly at the coincident
point, which satisfy,
\begin{align}
 \widetilde{\Box} Z^{ij}_{1(i(N+1))i'j'} &= -(N^2-(N-1)^2+2) Z^{ij}_{1(i(N+1))i'j'} \\
 \widetilde{\Box} Z^{ij}_{2(i(N+1))i'j'} &= -(N^2-N^2+2) Z^{ij}_{2(i(N+1))i'j'}
\end{align}
so we see that,
\begin{align}
 Z^{ij}_{1(i(N+1))i'j'}(l) &\propto W^{ij}_{(i(N-1))i'j'} (il) \label{N-1}\\
 Z^{ij}_{2(i(N+1))i'j'}(l) &\propto W^{ij}_{(iN)i'j'} (il) \label{N}
\end{align}
where $W^{ij}_{(p)i'j'}$ is defined in (\ref{tsumW}) and
can be written alternatively as in (\ref{defZpr}).
This is because the conditions mentioned are all that we used in obtaining
$W^{ij}_{(p)i'j'}$ in the first place.
(We have used $W^{ij}_{(p)i'j'}$ instead of $Z^{ij}_{(p)i'j'}$ here due to
the fact that $Z^{ij}_{(p)i'j'}$ may have poles for the values concerned.)
If indeed this is true for some non-zero proportionality constant,
this means that the derivatives of $E^{(i(N+1)v)}$ and $F^{(i(N+1)w)}_{i}$
give all the modes $\{ r'^{(i(N-1)u)}_{ij}\}$ and $\{ r'^{((iN)u)}_{ij} \}$ respectively.

The only potential problem lies in the fact that $Z^{ij}_{1(i(N+1))i'j'}$
and $Z^{ij}_{2(i(N+1))i'j'}$ might be zero.
From \cite{Allen:1985wd} and we see that,
\begin{align}
 Z_{(p)} &= C_p F(N+ip,N-ip;N+{1 \ov 2};1-z) \\
\begin{split}
 Z^i_{(p)i'} &= C'_p [\tilde{\gamma}^i_{i'}({2z(z-1) \ov N}{d \ov dz} + (2z-1) ) \\
 &+ n^i n_{i'}({2z(z-1) \ov N}{d \ov dz} + (2z -2)) ] \gamma_p(z) \\
 \text{for} &\quad \gamma_p(z) \equiv F(N+1+ip,N+1-ip;N+{3 \ov 2};1-z)
\end{split}
\end{align}
and from \cite{Camporesi:1994ga} we see that
\begin{align}
 C_p &\propto {[p^2+(N-1)^2] \Ga(ip+N-1) \Ga(-ip+N-1) \ov \Ga(ip) \Ga(-ip)} \\
 C'_p &\propto {[p^2+N^2] \Ga(ip+N-1) \Ga(-ip+N-1) \ov \Ga(ip) \Ga(-ip)}
\end{align}
up to a factor independent of $p$.
Although $C_p$ and $C'_p$ have poles, by direct calculation, we can obtain non-zero, non-sigular
$Z^{ij}_{1(i(N+1))i'j'}$ and $Z^{ij}_{2(i(N+1))i'j'}$.

Of course $Z^{ij}_{1(i(N+1))i'j'}$ and $Z^{ij}_{2(i(N+1))i'j'}$ can be obtained explicitly
to verify (\ref{N-1}) and (\ref{N}).



\begin{thebibliography}{99}

\bibitem{AdSCFT}
  J.~M.~Maldacena,
  Adv.\ Theor.\ Math.\ Phys.\  {\bf 2}, 231 (1998)
  [Int.\ J.\ Theor.\ Phys.\  {\bf 38}, 1113 (1999)]
  [arXiv:hep-th/9711200];
  O.~Aharony, S.~S.~Gubser, J.~M.~Maldacena, H.~Ooguri and Y.~Oz,
  Phys.\ Rept.\  {\bf 323}, 183 (2000)
  [arXiv:hep-th/9905111].

\bibitem{Strominger:2001pn}
  A.~Strominger,
  JHEP {\bf 0110}, 034 (2001)
  [arXiv:hep-th/0106113].

\bibitem{Dyson:2002nt}
  L.~Dyson, J.~Lindesay and L.~Susskind,
  JHEP {\bf 0208}, 045 (2002)
  [arXiv:hep-th/0202163].

\bibitem{Maldacena:2002vr}
  J.~M.~Maldacena,
  JHEP {\bf 0305}, 013 (2003)
  [arXiv:astro-ph/0210603].

\bibitem{Freivogel:2006xu}
  B.~Freivogel, Y.~Sekino, L.~Susskind and C.~P.~Yeh,
  Phys.\ Rev.\  D {\bf 74}, 086003 (2006)
  [arXiv:hep-th/0606204].

\bibitem{Susskind:2007pv}
  L.~Susskind,
  arXiv:0710.1129 [hep-th].

\bibitem{Bousso:2008as}
  R.~Bousso, B.~Freivogel, Y.~Sekino, S.~Shenker, L.~Susskind, I.~S.~Yang and C.~P.~Yeh,
  Phys.\ Rev.\  D {\bf 78}, 063538 (2008)
  [arXiv:0807.1947 [hep-th]].

\bibitem{Garriga:2008ks}
  J.~Garriga and A.~Vilenkin,
  arXiv:0809.4257 [hep-th].

\bibitem{ColemanVac}
  S.~R.~Coleman,
  Phys.\ Rev.\  D {\bf 15}, 2929 (1977)
  [Erratum-ibid.\  D {\bf 16}, 1248 (1977)];
  C.~G.~.~Callan and S.~R.~Coleman,
  Phys.\ Rev.\  D {\bf 16}, 1762 (1977).

\bibitem{Coleman:1980aw}
  S.~R.~Coleman and F.~De Luccia,
  Phys.\ Rev.\  D {\bf 21}, 3305 (1980).

\bibitem{Hertog:1999kg}
  T.~Hertog and N.~Turok,
  Phys.\ Rev.\  D {\bf 62}, 083514 (2000)
  [arXiv:astro-ph/9903075].

\bibitem{Gratton:1999ya}
  S.~Gratton and N.~Turok,
  Phys.\ Rev.\  D {\bf 60}, 123507 (1999)
  [arXiv:astro-ph/9902265].

\bibitem{Hawking:2000ee}
  S.~W.~Hawking, T.~Hertog and N.~Turok,
  Phys.\ Rev.\  D {\bf 62}, 063502 (2000)
  [arXiv:hep-th/0003016].

\bibitem{Allen:1985wd}
  B.~Allen and T.~Jacobson,
  Commun.\ Math.\ Phys.\  {\bf 103}, 669 (1986).

\bibitem{Camporesi:1994ga}
  R.~Camporesi and A.~Higuchi,
  J.\ Math.\ Phys.\  {\bf 35}, 4217 (1994).

\bibitem{Allen:1994yb}
  B.~Allen,
  Phys.\ Rev.\  D {\bf 51}, 5491 (1995)
  [arXiv:gr-qc/9411023].

\bibitem{Polishchuk:1999nh}
  A.~Polishchuk,
  JHEP {\bf 9907}, 007 (1999)
  [arXiv:hep-th/9905048].

\bibitem{Tanaka:1997kq}
  T.~Tanaka and M.~Sasaki,
  Prog.\ Theor.\ Phys.\  {\bf 97}, 243 (1997)
  [arXiv:astro-ph/9701053].

\bibitem{Erdmenger:1996yc}
  J.~Erdmenger and H.~Osborn,
  Nucl.\ Phys.\  B {\bf 483}, 431 (1997)
  [arXiv:hep-th/9605009].

\bibitem{Barth:1983hb}
  N.~H.~Barth and S.~M.~Christensen,
  Phys.\ Rev.\  D {\bf 28}, 1876 (1983).

\bibitem{D'Hoker:1999jc}
  E.~D'Hoker, D.~Z.~Freedman, S.~D.~Mathur, A.~Matusis and L.~Rastelli,
  Nucl.\ Phys.\  B {\bf 562}, 330 (1999)
  [arXiv:hep-th/9902042].

\end{thebibliography}
\end{document}